\documentclass{article}

\usepackage{arxiv}

\usepackage[utf8]{inputenc} 
\usepackage[T1]{fontenc}    
\usepackage{hyperref}       
\usepackage{url}            
\usepackage{booktabs}       
\usepackage{amsfonts}       
\usepackage{nicefrac}       
\usepackage{microtype}      
\usepackage{graphicx}
\usepackage{epstopdf, epsfig}
\usepackage{amsmath}
\usepackage{graphicx}
\usepackage{graphics}
\usepackage{subfig}
\usepackage{listings}

\usepackage{natbib}	
\usepackage{gensymb}
\usepackage{comment}
\usepackage{enumitem}
\setitemize{noitemsep,topsep=0pt,parsep=0pt,partopsep=0pt,leftmargin=*}
\usepackage{amssymb}

\usepackage{lineno}
\usepackage{setspace}
\usepackage{footnote}

\usepackage{xcolor}

\title{Bringing optical fluid motion analysis to the field: a methodology using an open source ROV as camera system and rising bubbles as tracers}
\author{
  Trygve K. Løken$^{a,}$\footnotemark[1], Thea J. Ellevold$^{a}$, Reyna G. Ramirez de la Torre$^{a}$,  Jean Rabault$^{b}$ and Atle Jensen$^{a}$
}

\begin{document}
\maketitle

\footnotetext[1]{E-mail address and phone number for corresponding: trygve.loken@gmail.com (+47) 94885767 (T.K. Løken)\\ 
E-mail addresses: theajel@math.uio.no (T. J. Ellevold), reynar@math.uio.no (R. De La Torre), \\ jeanr@met.no (J. Rabault) and atlej@math.uio.no (A. Jensen)}

\noindent $^{a}$ Department of Mathematics, University of Oslo, Oslo, Norway \\
$^{b}$ Norwegian Meteorological Institute, Oslo, Norway \\


\begin{abstract}

Detailed water kinematics are important for understanding atmosphere-ice-ocean energy transfer processes in the Arctic. There are few in situ observations of 2D velocity fields in the marginal ice zone. Particle tracking velocimetry and particle image velocimetry are well known laboratory techniques for measuring 2D velocity fields, but they usually rely on fragile equipment and pollutive plastic tracers. Therefore, in order to bring these methods to the field, we have developed a new system which combines a compact open-source remotely operated vehicle as an imaging device, and air bubbles as tracing particles. The data obtained can then be analyzed using image processing techniques tuned for field measurements in the polar regions. The properties of the generated bubbles, such as the relation between terminal velocity and diameter, have been investigated under controlled conditions. The accuracy and the spread of the velocity measurements have been quantified in a wave tank and compared with theoretical solutions. Horizontal velocity components under periodic waves were measured within the order of 10\% accuracy. The deviation from theoretical solutions is attributed to the bubble inertia due to the accelerated flow. We include an example from an Arctic field expedition where the system was deployed and successfully tested from an ice floe. This work is an important milestone towards performing detailed 2D flow measurements under the ice in the Arctic, which we
anticipate will help perform much needed direct observations of the dynamics happening under sea ice. 

\end{abstract}

\section{Introduction} \label{sec:Introduction}
The development of field studies of fluid flow in the ocean has increased considerably in the last decades thanks to the technological advances within measurement equipment. Nonetheless, working in the Arctic environment remains yet difficult. Not only because of the technical aspects to execute a successful controlled experiment in the field, but also because the possibilities to use regular laboratory equipment are limited by the harsh temperatures and conditions. On top of that, working on an ice layer brings new challenges to the design of a field setup. Adaptation of well known techniques to the field by accurately testing them in the laboratory first is a reasonable approach. This paper introduces a new system for high resolution velocity measurements of fluid flow specially designed for Arctic conditions, where elements from already established methods are combined. The accuracy of the instrumentation is determined to understand its capabilities in order to adapt them to the desired object of study.

Although laboratory experiments can yield useful results due to the high level of control over variables, scaling between wave tank conditions and the real ocean is challenging as several nondimensional numbers are necessary to describe the problem. In wave tank experiments, it is for example desirable to moderate nonlinear effects and ensure deep-water waves, as this is the most common situation in the Arctic. These conditions are achieved by respectively reducing the wave steepness and the ratio of wavelength over depth. As pointed out by \cite{rabault2019experiments}, it is with such constraints difficult to simultaneously obtain a realistic Reynolds number for turbulence under ocean waves. It is therefore necessary to perform field measurements to complement or confirm laboratory experiments, where scaling problems are usually inevitable.

Ultrasonic velocimeters, such as Acoustic Doppler Velocimeters (ADVs) and Acoustic Doppler Current Profilers (ADCPs) are well known instruments for fluid velocity measurements in the ocean. However, these instruments provide only single point velocity (ADVs), or in the best case an array of along-beam velocities (ADCPs), which in many cases is insufficient to give a detailed description of the dynamics of certain flow phenomena that occur in the Arctic. An example of such a phenomenon is wave propagation underneath an ice layer \citep{sutherland2016observations,rabault2017measurements}, which is interesting for validation of wave attenuation models through the marginal ice zone, e.g. \cite{weber1987wave,newyear1997comparison,sutherland2019two}. Another example is the investigation of turbulent structures or jets induced by colliding ice floes \citep{rabault2019experiments}, which may contribute to ice-ocean energy transfer and increased eddy viscosity and viscous dissipation. In these cases it is desirable to resolve a 2D velocity field for a better understanding of the underlying physics.

Within fluid dynamics and wave studies in particular, researchers have used two very established laboratory techniques to understand flow dynamics: Particle Image Velocimetry (PIV) and Particle Tracking Velocimetry (PTV). In controlled environments, these techniques can accurately represent the velocity fields in a wide range of flows. Conventional PIV and PTV systems normally utilize a powerful light source, such as lasers, to illuminate a thin sheet of tracer particles. In this way it can be ensured that the particles of interest are situated in a plane with a fixed distance to the camera. In later years it has been shown that these techniques can be applicable in field measurements with certain reserves, PIV-laser systems have been developed and deployed off docks and from ships \citep{bertuccioli1999submersible,smith2002piv}. However, it is difficult to arrange such a setup in the field, especially in an Arctic environment, due to the large weight, high cost and fragility of lasers.

Another complication in the use of traditional PIV and PTV in field measurements is the use of tracers. As addressed in \cite{bertuccioli1999submersible}, some oceanic areas can be analyzed by the use of the natural suspended particles. But, particularly in the Arctic Ocean, the water is mostly clear during the winter. Therefore, the introduction of environmental-friendly tracers is required. The adaptation of PIV using bubbles as tracers has become popular throughout different applications \citep{delnoij1999ensemble,de2015characterization,ryu2005use,rivillas2012estimation,cheng2005bubble}, and tested against more commonly seeding methods \citep{vargas2020estimation}. The mentioned studies have brought to light different aspects that need to be calibrated and considered to apply these techniques outside a controlled environment. In this study, an array of small air bubbles has been introduced as tracers, which provide a plane of particles and thus enables us to find a two-dimensional velocity field.

A good way to assist underwater measurements is the use of ROVs (Remotely Operated Vehicles). Since the early 1980s, the interest of utilizing ROVs for field work has increased. Towards the end of that decade, the first major field works using ROVs to retrieve information and samples were made \citep{ballard1993medea}. Afterwards, ROVs have been used extensively for observations and sample retrieving. By using an open source ROV, it is possible to control tools underwater and retrieve images and information from different sensors, while keeping costs down to a reasonable level. It is only in the last few years that ROVs have been considered a tool in different fluid dynamics applications and \textit{in situ} measurements, where the reliability and limitations of the sensors mounted remain the main issues \citep{pinkard2005use,keranen2012remotely}.

In this work, a methodology to perform reliable field measurements using PTV and PIV techniques is described. Extensive laboratory tests have been included to show the reliability and range of confidence of the bubble curtain as a substitute for tracers and laser sheets. The use of an ROV as an imaging and light source system in PTV/PIV field measurements has been introduced and tested in extreme Arctic conditions. For simplicity, we introduce the abbreviation PV (Particle Velocimetry) to address PTV/PIV and name the introduced technique "ROV-PV system". The paper is organized in the following manner. Section~\ref{sec:Data_methods} describes the components and the assembly of the ROV-PV system, the algorithms used for image processing and how fluid velocity is interpreted from measured bubble velocity. Section~\ref{sec:Wave_velocity} presents the results from the laboratory experiments. An example of a field deployment in the Arctic is included in Section~\ref{sec:Field_deployment}. The methods and results are discussed in Section~\ref{sec:Discussion} and conclusions are drawn in Section~\ref{sec:Conclusions}.

\section{Data and methods} \label{sec:Data_methods}

The ROV-PV system presented here is designed to capture dynamics generated from air-water and ice-water interactions in the upper ocean. It is a portable system build for Arctic field campaigns at remote locations and harsh environments. The core concept is to perform optical recordings with the camera of an ROV. A plane of rising air bubbles illuminated by the ROV's headlights are used as tracing particles. The bubbles rise inside a thin aluminum frame with indicated reference coordinates. Images are calibrated to obtain real world units. Image processing is applied to produce the 2D in-plane velocity field of the bubbles. During post processing, the vertical buoyancy driven velocity component of the bubbles is subtracted to obtain the water velocity.

\subsection{ROV-PTV system} \label{subsec:ROV}

An open source ROV has been chosen for image acquisition that will be used for the PV analysis, as it is easily maneuvered below and around sea ice. The use of open source instruments is an increasing trend in geophysics, see e.g. \cite{lockridge2016development,rabault2016measurements,rabault2020open}. Open source instruments provide flexibility in the sense that they offer the possibilities for extensive modification to specific needs and easy interfacing with other open source systems. Another advantage is the significantly lower cost than similar closed source solutions. An ROV allows for preliminary inspections of the site to find the most suitable location for measurements, as opposed to a stationary camera. The high-performance BlueROV2 from BlueRobotics \citep{BlueRobotics} with open source software and electronics has been used. It is installed with eight thrusters to increase the stability and obtain full six degrees of freedom control. The ROV is physically connected to a field computer by a tether, which carries and transfers video and data signal. An open source application called QGroundControl (QGC) works as the user interface and provides live video stream and various types of information for the pilot.

The camera is a wide-angle low-light HD USB camera with 1920$\times$1080 pixel resolution. It is mounted on a servo motor for user control of tilt angle. The camera lens has a 2.97~mm focal length with low distortion (1\%) and a 80\degree horizontal and 64\degree vertical Field Of View (FOV). The highest available frame rate $\Delta f =$~30~frames/s is used for maximum temporal resolution. Camera settings such as exposure, brightness and gain are adjusted in QGC to optimize the images in different environments and conditions, to ensure that the bubbles appear as clear, circular particles and not as blurry streaks. The ROV is equipped with four controllable headlights with a total capacity of 6000 lumens.

\begin{figure}
  \begin{center}
    \includegraphics[width=.55\textwidth]{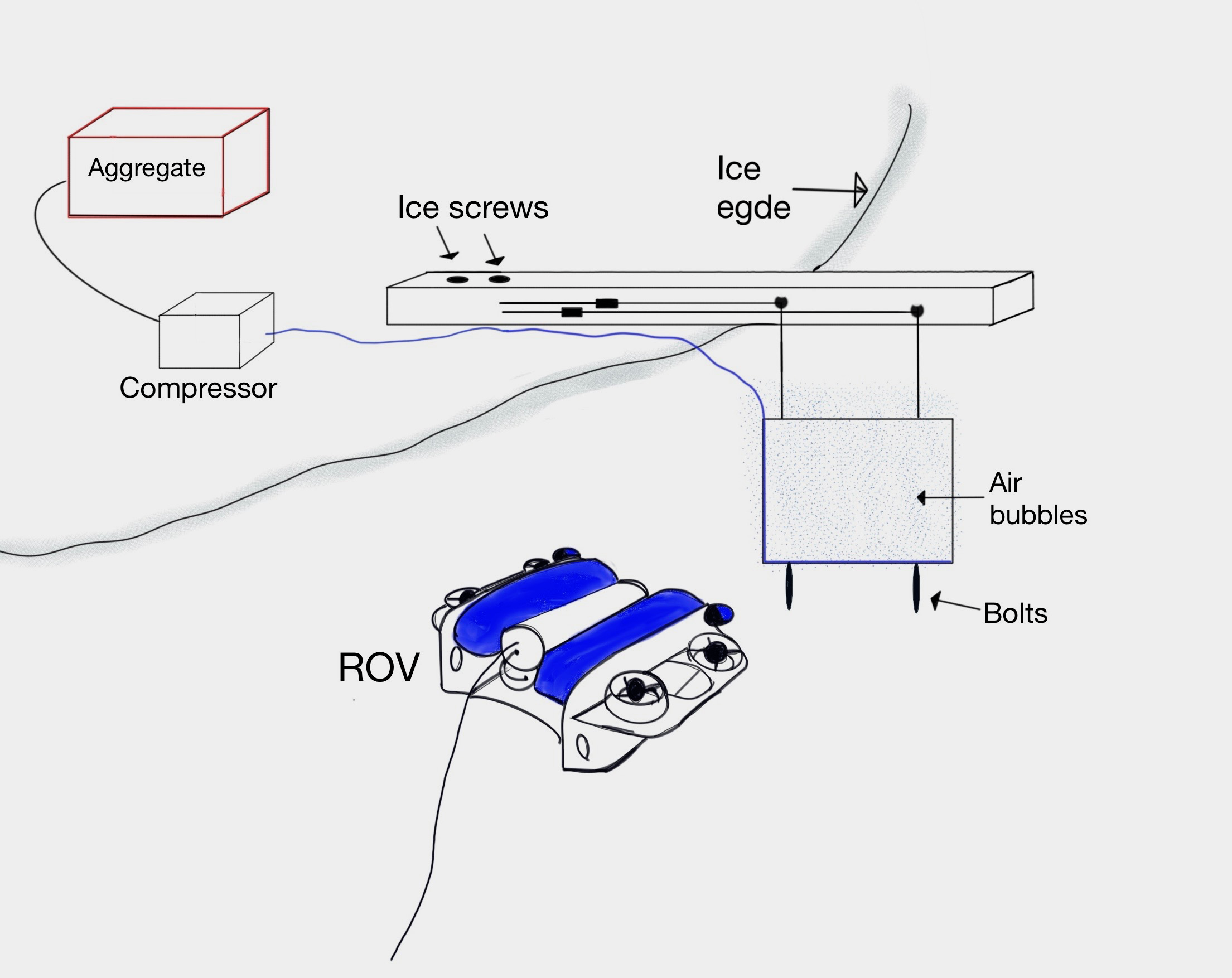}
    \caption{\label{fig:ROV-PTV_setup} Schematic of the ROV-PV system in an Arctic environment. The 2D in-plane water velocity is estimated from the bubble motion, which is recorded with the ROV camera. The perforated hose can for example be suspended into the water by means of a pulley system, as shown here, or directly through a hole in the ice. }
  \end{center}
\end{figure}

In the setup, which is illustrated in Fig.~\ref{fig:ROV-PTV_setup}, bubbles are generated with a 5~m long flexible silicon rubber hose. At the one end, the hose is fed with air of approximately 0.5~bar from a 1.1~kW ABAC compressor. At the other end, the last 0.75~m is perforated every 1~cm on the upward facing side with a 0.3~mm needle and the tube is sealed with a plug to prevent air leakage. This configuration provides an array of relatively small bubbles with a diameter of 0.6$-$1.4~mm. Bubbles have also been generated with a thin carbon fiber pipe perforated with a 0.1~mm drill. This device was not used in the experiments because the bubbles grew too large while sticking to the pipe before they detached. The headlights of the ROV have proven sufficient to make the bubbles visible during fieldwork. During laboratory experiments, the ROV has been placed outside of the water tank due to space limitations, and the bubbles have been illuminated with external LED lamps to avoid headlight reflections in the glass wall.

In order to obtain quantitative physical information from the experiments, it is necessary to implement a coordinate system to be able to convert from pixels to real-world units in the post processing. The coordinate system needs to be in the framework of the bubble plane. A thin aluminum grid with a combination of grid bars and woven string has been used as coordinate indicators. The grid measures 55$\times$45~cm and the coordinate resolution is 8$-$12~cm in the horizontal and 9~cm in the vertical direction. In a traditional particle tracking setup, the camera and the light-sheet are fixed in one position throughout the whole experiment, making it easy to take a picture of a coordinate system in advance of the measurements. In the current system, the ROV will always be in some movement which leads to a varying position and angle relative to the coordinate grid. Therefore, the perforated hose is integrated at the bottom part of the grid so that each single image can be converted into real-world units with a third order coordinate transformation, e.g. \cite{kolaas2016getting}. Upon deployment in the ocean, the perforated hose and coordinate system are suspended from a constructed frame or from the sea ice. Heavy bolts are attached to the bottom of the grid to keep it vertical and in place when exposed to waves and ocean currents.

\subsection{PTV and PIV methodology} \label{subsec:PTV_methodology}

In this work, both PTV and PIV were used in different cases. PTV differs from PIV in a fundamental way. PIV relies on pattern matching in an essentially Eulerian way, whereas PTV seeks to identify individual particles and follow them in a Lagrangian sense. 
Therefore, it is important to choose the most suitable technique based on the characteristics of the experiment, and on the information which is desirable to obtain. A small review of the PTV and PIV algorithms used for this work follows.

DigiFlow Software \citep{dalziel2006digiflow} has been utilized for PTV in this study. After the image acquisition, the image is scanned for blobs that have an intensity satisfying some threshold parameters. If a blob is found, its characteristics are determined and compared with a set of requirements for the blob to be considered a particle. If the blob satisfies these requirements, it is recorded as a particle, if it does not, it is discarded. 
Once all the particles in an image have been found at $t = t_n+1$, they need to be related back to the previous image $t = t_n$. DigiFlow uses a modification of what is known in operations research as the Transportation Algorithm, which was developed by \cite{dalziel1992decay}. 
The algorithm chooses a set of particles $P$ at $t = t_n$, and the other the set of particles $Q$ at $t = t_n+1$. At $t = t_n$, $p_i$ for $i=1,2,...,M$ represents the set of particles, while at $t = t_n+1$, they are labeled $q_j$ for $j=1,2,...,N$. Each $p_i$ or $q_j$ contains the location of the particle and other characteristics such as size, shape, intensity, or any other desired information. A set of association variables $\alpha_{ij}$ is defined. When $\alpha_{ij} = 1$  then $p_i$ at $t = t_n$ is the same particle as $q_j$ at $t = t_n+1$. If $\alpha_{ij} = 0$, then $p_i$ and $q_j$ represent different physical particles. The number of particles in the images may be different at $t = t_n$ and $t = t_n+1$. To overcome this problem, $\alpha_{0j}$ and $\alpha_{i0}$ are defined as dummy particles at times $t = t_n$ and $t = t_n+1$, respectively. Unlike ordinary particles, more than one value of $j$ or $i$ may give a nonzero value of $\alpha_{0j}$ and $\alpha_{i0}$, respectively. In this case, a nonzero value of $\alpha_{0j}$ indicates that particle $p_i$ at $t = t_n$ has been lost from the image by $t = t_n+1$, either by moving out of the image or for some other reason. Similarly, $\alpha_{0j} = 1$ represents a particle $q_j$ present at $t = t_n+1$ which was not there at $t = t_n$.

The in-house HydrolabPIV software developed at the University of Oslo has been used for PIV in this work \citep{kolaas2016getting}. This software combines different techniques presented in previous works to optimize the PIV algorithm \citep{adrian1991particle,westerweel2005universal,padfield2011masked,raffel2018particle}. Given two consecutive images of a seeded flow, it is desirable to find an Eulerian description of the velocity field. To achieve this, the images are divided into a regular grid of \textit{subwindows}, usually of size from 8$\times$8 to 64$\times$64~pixels. The use of a certain percentage of overlap is also common, usually between 25\% and 75\%. For each subwindow in the first frame $I_1$, a subwindow $I_2$ with a similar pattern is searched in the second frame. A metric of the similarity of the patterns is required, a common choice is the normalized cross-correlation

\begin{equation}
\label{eq:PIV_cc}
R_{ncc} = \frac{\sum I_1 I_2 - \frac{\sum I_1 I_2}{N}}{\sqrt{\left( \sum I_1^2 - \frac{(\sum I_1)^2}{N} \right) \left(  \sum I_2^2 - \frac{(\sum I_2)^2}{N} \right)}}.
\end{equation}

To speed up the processing, fast Fourier transform is often used to calculate the cross-correlation. An ensemble averaged velocity is found by dividing the optimal match displacement of the pattern by $\Delta t $. After the velocity is found, the result can be made more robust by checking the quality of the image, and further validation of the vectors using outlier detection. This detection uses a 3$\times$3 normalized local median filter. In addition, a cubic B-spline is fitted to the velocity field using an iterative weighted least squares fit. The local residuals are used with the biweight function in the first iteration. The fit is then iterated a few times where the residuals are used to update the weights in the least squares fit. The normalized residuals from the last iteration together with the residuals from the local median filter are used to mark the outliers. These outliers can be removed and replaced by using the fitted B-spline.

\subsection{Air bubbles as tracers} \label{subsec:Bubbles}

Bubbles as tracing particles are challenging because they do not passively follow the fluid motion as opposed to conventional tracers. The buoyancy driven motion of the bubbles must be subtracted in order to achieve the absolute water velocity. In the following, the $\mathbf{x} = (x,y)$, $\mathbf{V} = (u,v)$ and $\mathbf{F} = (F_{x},F_{y})$ conventions will be used for 2D position, velocity and force, respectively. The objective of this study is to introduce a technique where the velocity of bubbles $\mathbf{V_{b}}$ is measured in order to determine the velocity of the water $\mathbf{V_{w}}$. An important quantity in this method is the slip velocity $\mathbf{V_{slip}}$ which is the relative velocity between the water and the bubbles, defined as 

\begin{equation}
\label{eq:Relative_vel}
\mathbf{V_{slip}} = \mathbf{V_{b}}-\mathbf{V_{w}}.
\end{equation}

Small bubbles are beneficial because they have rectilinear vertical trajectories, whereas larger bubbles tend to oscillate in the horizontal direction \citep{haberman1953experimental}, which raises the need for further velocity corrections. It is additionally advantageous to generate uniform bubbles along the array, which will allow for subtraction of the mean buoyant velocity across the entire velocity field. As pointed out by \cite{oguz1993dynamics}, it is difficult to produce small bubbles of uniform size and rate. In the present study, the variability in bubble diameter was large over the span of the perforated hose. From visual inspection, the bubble diameter seemed to be randomly distributed in size. It is likely that the difference in bubble size was caused by the non-uniform size of the outlet holes, since the silicon rubber material may have contracted differently from hole to hole after the needle was subtracted.




For steady uniform water flow, bubbles will approximately follow the water motion in the horizontal direction after a relaxation time. An additional vertical component is present due to the buoyancy force $\mathbf{B} = \frac{4}{3}(\rho_w - \rho_b)g\pi R^{3}$, where $\rho_{w}$ denotes the water density, $\rho_{b}$ the bubble density, $g$ the acceleration due to gravity and $R$ the bubble radius. In this case, $\mathbf{V_{slip}} = v_{slip}$ only has a vertical component driven by $\mathbf{B} = B_{y}$. On the other hand, $\mathbf{V_{slip}}$ also depends on the bubble inertia when the water flow $\mathbf{V_{w}}(\mathbf{x},t)$, where $t$ denotes time, is accelerated. Since $\frac{\rho_{b}}{\rho_{w}} \ll 1$, the effect of added mass becomes important to determine the inertia. Drag force on a spherical bubble is expressed as $\mathbf{D}= \frac{1}{2}\rho_{w}\pi R^{2} C_{D} \mathbf{V_{slip}}|\mathbf{V_{slip}}|$. The drag coefficient $C_{D}$ depends on the Reynolds number $Re=\frac{2R|\mathbf{V_{slip}}|}{\nu}$, where $\nu$ denotes the water kinematic viscosity. Following \cite{grue1999properties}, an estimate of the slip velocity can be obtained from momentum balance of a bubble 


\begin{equation}
\label{eq:Momentum_balance}
\mathbf{D(V_{slip})} - \mathbf{B} = M_{b}\left[\frac{\partial \mathbf{V_{w}}}{\partial t}\right]_{max},
\end{equation}

\noindent where $M_{b} = \rho_{w}\frac{2}{3}\pi R^{3}$ is the added mass of a spherical bubble, and the mass of the bubble is neglected. The expression for $\mathbf{V_{slip}}$ depends on the relation between $C_{D}$ and $Re$ and on the maximum water acceleration $[\partial \mathbf{V_{w}}/\partial t]_{max}$. In Section~\ref{sec:Wave_velocity}, the horizontal velocity component under the crests of water waves, where there exist analytical approximations for the acceleration, is investigated. An analytical relation for $C_{D}(Re)$ is used to solve Eq.~(\ref{eq:Momentum_balance}) for $\mathbf{V_{slip}}$ and it is shown that the observations are in agreement with the error analysis. 

Analytical solutions exist for $C_{D}(Re)$ for spherical bubbles moving in infinite mediums at low Reynolds numbers. Stokes' solution, i.e. $C_{D} = \frac{24}{Re}$ is a good approximation for $Re < 1$ \citep{haberman1953experimental}. For moderate Reynolds numbers, boundary-layer theory approximations have resulted in analytical solutions. \cite{batchelor2000introduction} assumed an irrotational flow outside a thin boundary layer around spherical bubbles and use potential theory to obtain $C_{D} = \frac{48}{Re}$, which is a fair estimate for 20~$< Re < $~200$-$500, depending on the fluid. Bubbles tend to stay spherical up to and including moderate Reynolds numbers due to sufficiently strong surface tension forces. In the spherical regime, the bubble trajectory is usually rectilinear and the vertical velocity increases with the radius squared.  Above $Re \approx$~450, bubbles enter the elliptical regime where they start to deform and flatten as viscous and hydrodynamic forces become more prominent. Consequently, the drag increases, and approximations of $C_{D}$ primarily rely on numerical simulations and experimental data. In the ellipsoidal regime, the rising motion is typically helical or oscillatory and the terminal velocity has been observed to increase more slowly or even decrease with radius \citep{clift2005bubbles}. \cite{haberman1953experimental} reported a smooth transition in vertical velocity between the spherical and the elliptical regime in fresh water. For $Re > $~5000, bubbles normally take form as spherical caps, but this region is not relevant for the present study. 


The terminal velocity $v_{T}$ of a bubble can be found theoretically by balancing the drag and the buoyancy force in the vertical direction. For a comparison with the observed bubble terminal velocity presented next, the analytical relation for $C_{D}(Re)$ of \cite{batchelor2000introduction} is used, i.e. $C_{D} = \frac{48}{Re}$. This gives a drag force $D_{y} = 12\mu_{w} \pi R v_{T}$. The force balance is rearranged and solved for the terminal velocity 

\begin{equation} \label{eq:Terminal_vel}
	v_{T} = \frac{(\rho_w - \rho_b)gR^2}{9\mu_w}.
\end{equation}



\begin{figure}
        \centering
        \includegraphics[width=.55\textwidth]{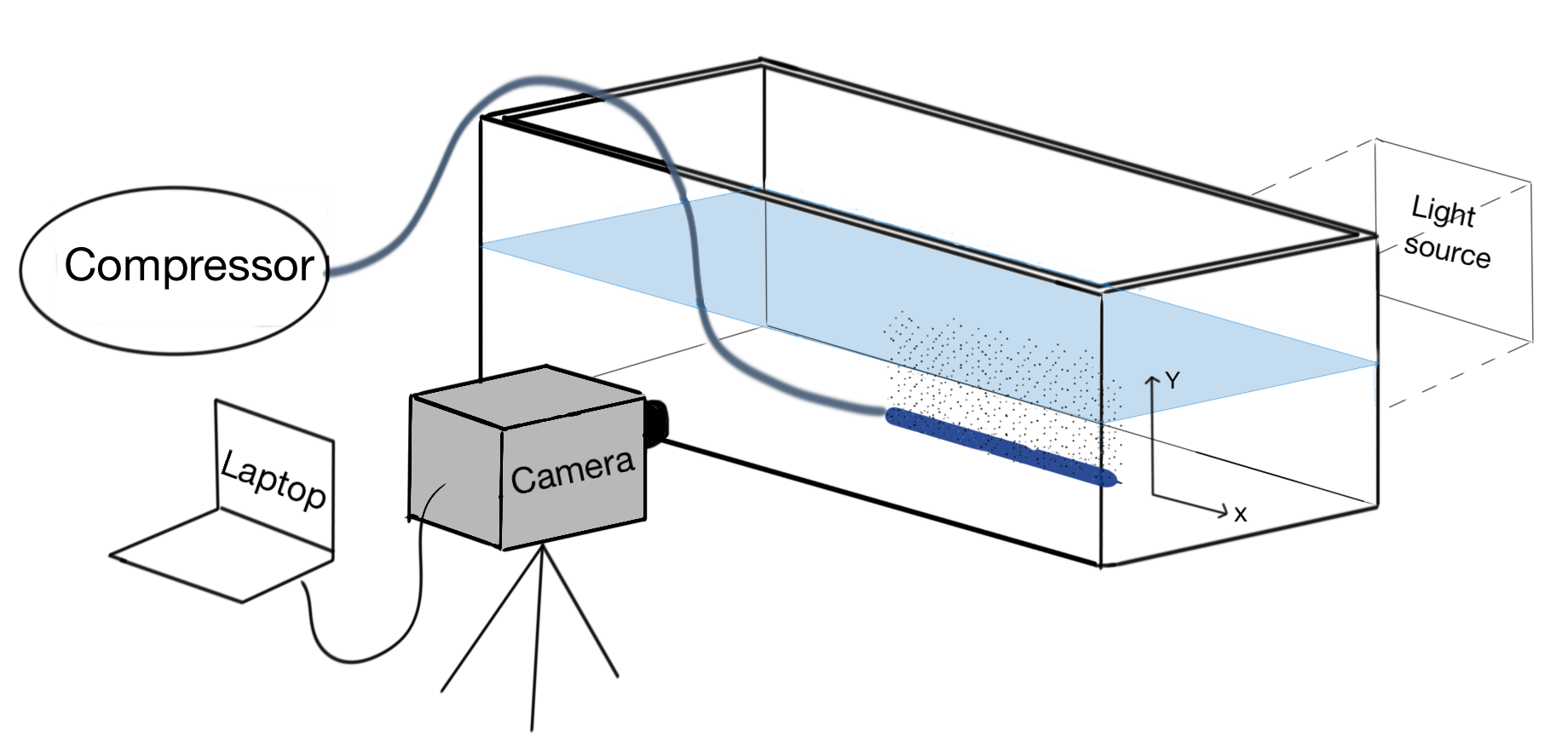}	
        \caption{ Schematic of the experimental setup to determine the terminal velocity of rising bubbles in stagnant fluids with $(x, y)$ axis aligned horizontally and vertically, respectively. The bubble motion was recorded with a high speed camera and the bubble velocity and size were found from image processing techniques.}
        \label{fig:SmallTank}
\end{figure}

The relation between the size and the rising velocity of bubbles produced by the perforated hose has been investigated in the Hydrodynamical Laboratory at the University of Oslo. The experiments were performed in a small glass tank measuring 1.5~m long and 0.4~m wide with water depth $h=$~27~cm. The perforated hose was placed on the bottom, parallel to the long wall. In this experiment, the motion of the rising bubbles was of interest, and not the ROV-PV system as a whole. Therefore, the ROV camera was substituted with a Photron FASTCAM high speed camera for increased accuracy. Fresh water was used in the first case and salt water with salinity equal to typical Arctic conditions was used in the second case. A LED lamp placed on the back side of the tank illuminated the bubbles. An integrated laboratory Kaeser compressor produced a stable airflow to the perforated hose. A schematic of the experimental setup is shown in Fig.~\ref{fig:SmallTank}. The water temperature was 18$^{\circ}$C. Other fluid parameters are summarized in Table~\ref{tbl:calc_exp}. 

\begingroup
\setlength{\tabcolsep}{6pt} 
\renewcommand{\arraystretch}{1.2}
\begin{table}[!h]
\centering
\begin{tabular}{l  l  l} \toprule
Parameter &  Fresh water ($f$) & Salt water ($s$) \\ \midrule
Density $[kg/m^3]$ & $\rho_w = 998$ & $\rho_w = 1026$ \\
Dynamic viscosity $[Pa$ $s]$ & $\mu_w = 1.027 \times 10^{-3}$ & $\mu_w= 1.103 \times 10^{-3}$ \\
\bottomrule
\
\end{tabular}
\caption{Parameters used in the experiments. These values are used for calculating the theoretical terminal velocity in both fresh and salt water.} 
\label{tbl:calc_exp}
\end{table}
\endgroup

The image resolution was 1024$\times$1024 pixels. The frame rate and shutter speed were set to 500~frames/s and 1/9000~s, respectively. The camera was located outside the tank, 15~cm from the tank wall. The FOV for the validation was 40.1$\times$40.1~mm in fresh water and 48.8$\times$48.8~mm in salt water. To be certain that the terminal velocity was reached, the center of FOV was placed 18~cm above the tank bottom. The vertical displacement of a bubble between two frames varied 5$-$20~pixels, depending on the bubble speed.

Throughout the experiments, bubbles interacted with each other and in some situations, the tracking algorithm mistook two bubbles for one. These bubbles were filtered out. Additionally, bubbles which entered an elliptical regime were removed in order to compare the results with theoretical solutions for spherical bubbles. Elliptical bubbles and misinterpretations were filtered out by applying a ratio parameter $\beta = \frac{\text{vertical radius}}{\text{horizontal radius}}$. $\beta$ contains information on the bubble shape, e.g. $\beta \approx$~1 implies a spherical shape and $\beta <$~1 implies a horizontal elliptic shape. Table~\ref{tbl:thresholds1} lists the threshold values of $\beta$ used in the validation. The horizontal and vertical bubble diameters were determined with 4~pixel accuracy, which corresponds to an error up to approximately 14\%.

The instantaneous horizontal ($u_{i}$) and vertical ($v_{i}$) bubble velocities were calculated from the position ($x_{i},y_{i}$) for each frame $i$ a bubble was tracked. Here, $i = 1,2,...,N-1$, where $N$ is the number of frames, typically 50$-$200 depending on the bubble speed. The minimum, mean and maximum horizontal and vertical velocity were found for each bubble. The standard deviation of vertical velocity ($\sigma_v$) and horizontal radius ($\sigma_r$) for each bubble were used as indicators for consistency, and bubbles which did not satisfy the quality parameters listed in Table~\ref{tbl:thresholds1} were not included in the analysis. The bubbles included in the results were very consistent with $\sigma_v$ and $\sigma_r$ both in the order of $10^{-3}$ in fresh and salt water. This indicates that the vertical velocity was quite constant and that the terminal velocity was reached. 

\begin{table}[h!]
\center
\begin{tabular}{ r c c} \toprule
Parameter & Fresh water &  Salt water\\ \midrule
$Ratio$ & $0.6 \leq \beta \leq 1.0 $ & $ 0.6 \leq \beta \leq 1.0$  \\
$\sigma_r$ $[mm]$ & $\sigma_r \leq 0.012$ & $\sigma_r \leq 0.015$ \\
$\sigma_v$ $[ms^{-1}]$ & $\sigma_v \leq 0.007$ & $\sigma_v \leq 0.01$ \\ \bottomrule
\
\end{tabular}
\caption{Threshold parameters applied when processing the data from the bubble terminal velocity experiments.} 
\label{tbl:thresholds1}
\end{table}

\begin{figure}[!h]
       \centering
       \includegraphics[width=0.49\linewidth]{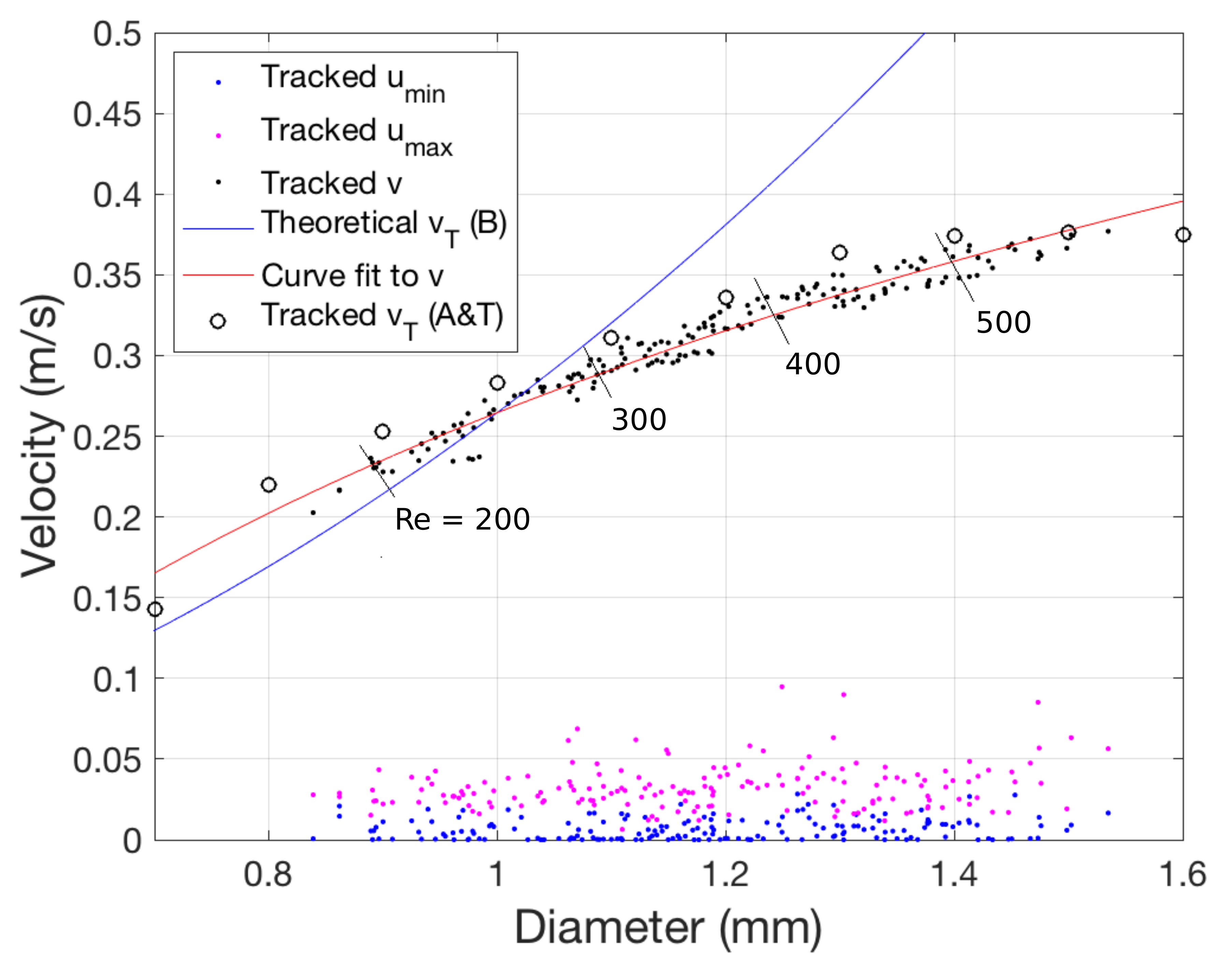}
       \includegraphics[width=0.49\linewidth]{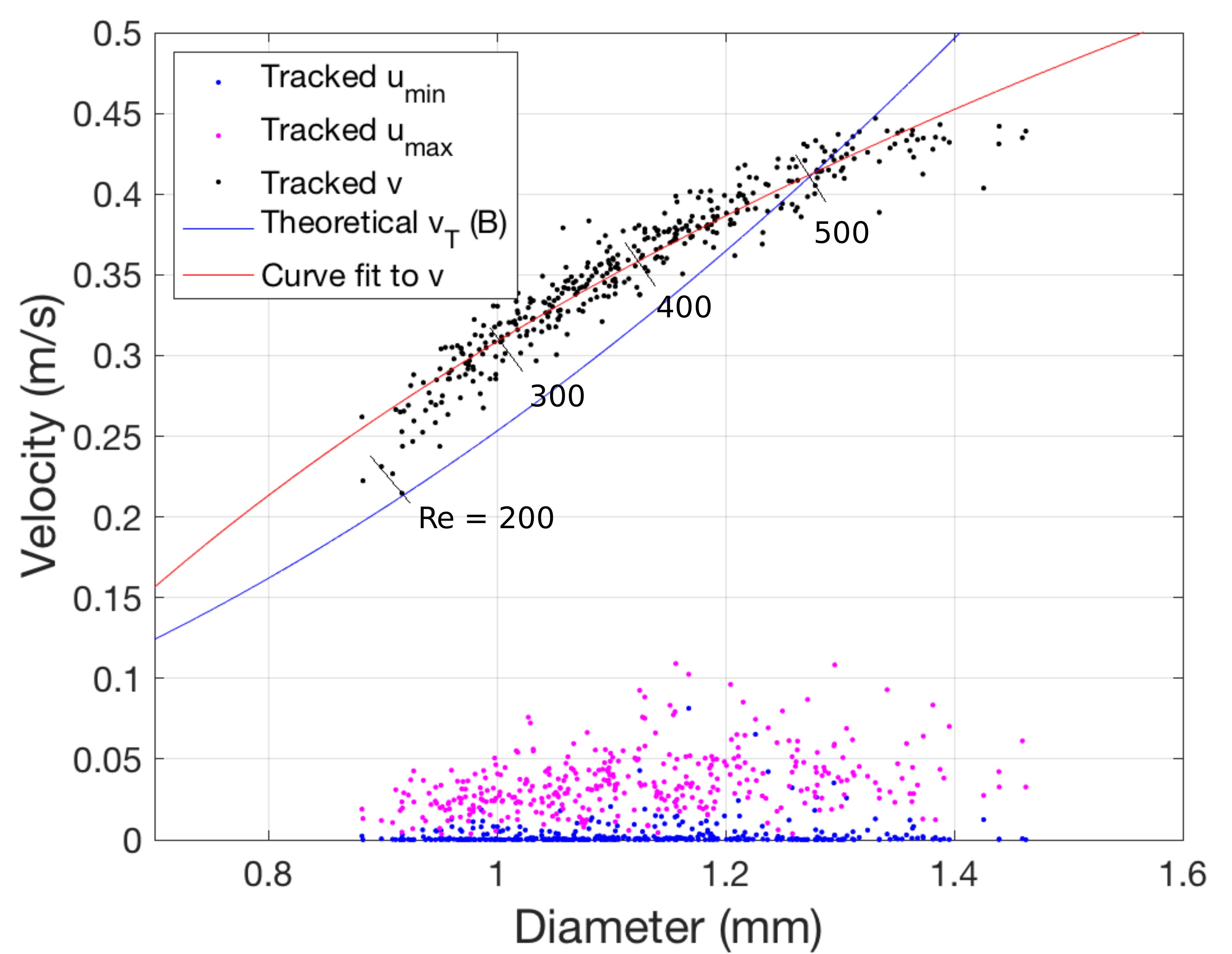}
      \caption{Measured bubble terminal velocity vs diameter (black dots) from experiments in fresh water (left panel) and salt water (right panel) at 18$^{\circ}$C. Only bubbles satisfying the quality criteria in Table~\ref{tbl:thresholds1} are included. The theoretical model of \cite{batchelor2000introduction} (blue line) is a fair estimate up to $Re \approx$~300 for fresh water and $Re \approx$~500 for salt water. Reynolds numbers of the tracked bubbles are indicated. The experimental results of \cite{aybers1969motion} (Fig. 15, curve I) are included for comparison with the fresh water experiment (black circles) and agree well with the current observations. A logarithmic curve fit has been applied to the tracked terminal velocity (red line), and is used in Section~\ref{sec:Wave_velocity} as a relation between bubble terminal velocity and diameter. Blue dots show minimum and purple dots show maximum tracked absolute horizontal velocity. }
      \label{fig:diam_vel_oldtube}
\end{figure}

Figure~\ref{fig:diam_vel_oldtube} presents the tracked terminal velocity of the rising bubbles (black dots) in fresh water (left panel) and salt water (right panel). A logarithmic increase in terminal velocity with respect to the bubble diameter can be observed. A logarithmic curve fit has been applied to the terminal velocity in both stagnant fluids by means of linear least squares (red line). Tracked maximum and minimum absolute horizontal velocities are displayed with purple and blue dots, respectively. There is a slight increase in maximum absolute horizontal velocity with respect to bubble diameter in the case of fresh water. This increase is more prominent in salt water. The theoretical terminal velocities calculated from Eq.~(\ref{eq:Terminal_vel}) with values listed in Table~\ref{tbl:calc_exp} and $\rho_b = $~1.2~kg/m$^3$ are displayed as blue lines.  

\cite{aybers1969motion} conducted a similar study on the terminal velocity of rising bubbles in stagnant water. Their results (Fig. 15, curve I), included in Fig.~\ref{fig:diam_vel_oldtube} (left panel) as black circles for comparison, are in agreement with the present findings. \cite{aybers1969motion} found that the bubbles entered the ellipsoidal regime for $Re \approx$~530, where the terminal velocity started to decrease. Whether the present bubbles enter the ellipsoidal regime in fresh water is not clear from visual inspections, the change in regime may occur at a higher Reynolds number than investigated here. The salt water bubbles possibly enter the ellipsoidal regime above $Re \approx$~540, where we observe the terminal velocity to flatten out. The theoretical terminal velocity from Eq.~(\ref{eq:Terminal_vel}) is a fair estimate up to $Re \approx$~300 for fresh water and $Re \approx$~500 for salt water, although the curve shape is quite different, most likely because the bubbles investigated are in the transition between the spherical and ellipsoidal regime. 

\section{Validation in a wave tank} \label{sec:Wave_velocity}

The ROV-PV system was evaluated under propagating periodic surface waves. A comparison to theoretical solutions to determine the accuracy of the measurements follows. The experiments were carried out in a wave tank in the Hydrodynamical Laboratory at the University of Oslo.  


Plane progressive waves are characterized by the acceleration of gravity $g$, the angular wave frequency $\omega$, the amplitude $a$, the wave number $k$ and the corresponding wavelength $\lambda$, where $\omega = 2\pi f$ and $\lambda = 2\pi / k$. In the case of $h > \frac{1}{2}\lambda$, the effects of the bottom are negligible. Periodic waves with steepness $ak \ll$~1 can be well approximated by third order Stokes theory, where the velocity potential $\phi$ is given by 


\begin{equation}
\label{eq:Potential_deep}
\phi = \frac{a g}{\omega}e^{k y}sin(k x-\omega t)+O(a^{4}),
\end{equation}       

\noindent where $x$ denotes the horizontal axis pointing in the direction of wave propagation and $y$ the vertical axis pointing upwards with $y =$~0 as the undisturbed free surface \citep{newman2018marine}. Equation (\ref{eq:Potential_deep}) is accurate up to and including terms of $O(a^{3})$ as long as the second order dispersion relation

\begin{equation}
\label{eq:Dispersion}
\omega^{2} = gk(1+a^{2}k^{2})+O(a^{3}k^{3}),
\end{equation}

\noindent which describes the relation between $\omega$ and $k$, is applied \citep{newman2018marine}. From Eq. (\ref{eq:Potential_deep}), the following terms for horizontal and vertical velocity components, respectively, are obtained up to and including third order

\begin{equation}
\label{eq:Velocity_liquid_u}
u_{w} = \frac{\partial \phi}{\partial x} = \frac{g a k}{\omega} e^{k y}cos(k x-\omega t),
\end{equation} 

\begin{equation}
\label{eq:Velocity_liquid_v}
v_{w} = \frac{\partial \phi}{\partial y} = \frac{g a k}{\omega} e^{k y}sin(k x-\omega t).
\end{equation}

Following \cite{jensen2001accelerations}, the velocities are nondimensionalized with $g a k/\omega$ for the comparison of waves with different amplitudes. The velocities in vertical columns directly underneath the wave crests are investigated in the further analysis. Here, the phase functions are $k x-\omega t = 2\pi n$, where $n$ is any integer. The nondimensional form of Eqs. (\ref{eq:Velocity_liquid_u})$-$(\ref{eq:Velocity_liquid_v}) evaluated under the wave crest reduce to $u_{w}/(g a k/\omega) = e^{k y}$ and $v_{w}/(g a k/\omega) = 0$, respectively. In the following, the non-zero horizontal velocity component is investigated.

In Section~\ref{subsec:Bubbles}, it was shown from the momentum equation that the slip velocity is related to the bubble inertia due to added mass. Here, it is attempted to estimate $u_{slip}$ by evaluating the momentum equation in the horizontal direction. As a wave undergoes an entire period, the maximum horizontal water acceleration up to and including third order terms is       

 \begin{equation}
\label{eq:Acceleration_liquid}
\left[\frac{\partial u_{w}}{\partial t}\right]_{max} = g a k e^{k y}.
\end{equation}

It is assumed that viscosity dominates the effect of inertia of the fluid in the horizontal direction, due to the small bubble size and horizontal slip velocity. Therefore, the drag force $D_{x}$ acting on the bubbles is approximated with the Stokes drag, i.e. $D_{x} = 6\rho_{w}\nu \pi R u_{slip}$. By applying the Stokes drag approximation, a linear relation between the drag force and the slip velocity is obtained. It should therefore be reasonable to perform a decomposition of the drag force into the vertical and horizontal direction. Equation (\ref{eq:Acceleration_liquid}) and the drag force approximation are inserted into Eq. (\ref{eq:Momentum_balance}), which is rearranged to estimate the slip velocity 

 \begin{equation}
\label{eq:slip_vel}
u_{slip} \approx \frac{R^{2}}{9\nu}\left[\frac{\partial u_{w}}{\partial t}\right]_{max} = \frac{R^{2}}{9\nu} g a k e^{k y},
\end{equation}

\noindent where the buoyancy term is neglected, as we are only looking at the horizontal component. Equation (\ref{eq:Velocity_liquid_u}) and (\ref{eq:slip_vel}) are inserted into Eq. (\ref{eq:Relative_vel}) to obtain an estimate for the horizontal bubble velocity 

 \begin{equation}
\label{eq:bubble_vel}
u_{b} \approx \frac{g a k}{\omega} e^{k y}\left(1-\frac{\omega R^{2}}{9\nu}\right).
\end{equation}
   

\begin{figure}
  \begin{center}
    \includegraphics[width=.55\textwidth]{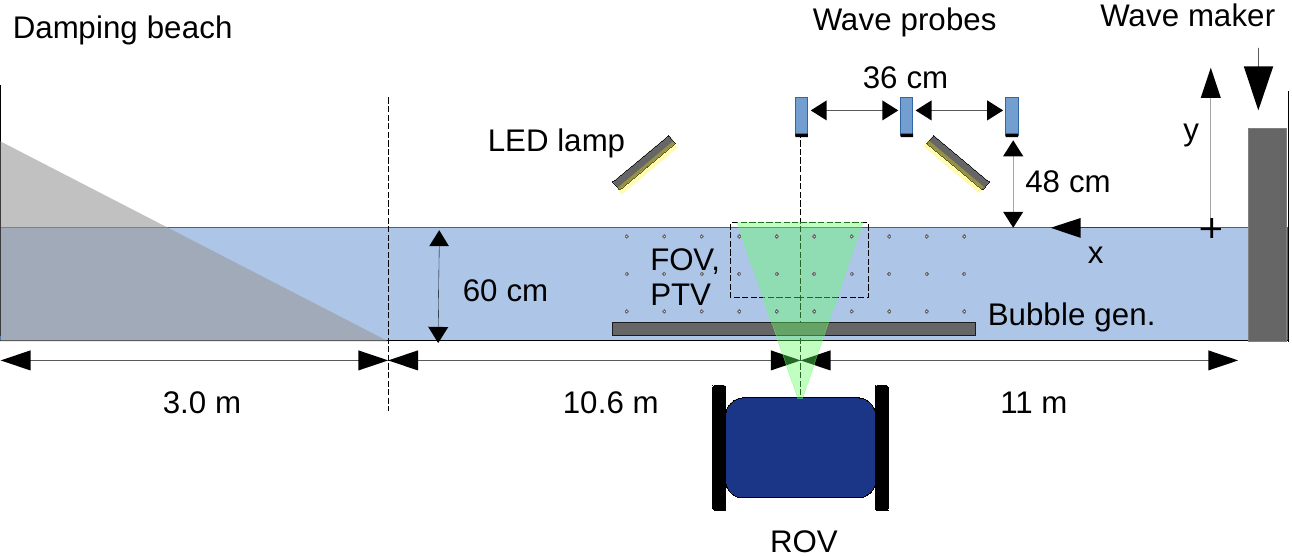}
    \caption{\label{fig:wave_setup} Schematic drawing of the experimental setup in the wave tank. A coordinate system is defined with the $(x, y)$ axes aligned horizontally in the direction of wave propagation and vertically in upward direction, respectively, with $y =$~0 as the undisturbed free surface. The velocity under the waves is estimated from the motion of the bubbles, which is recorded with the ROV camera.}
  \end{center}
\end{figure}

Figure \ref{fig:wave_setup} shows a schematic drawing of the experimental setup in the wave tank. It measures 24.6~m long and 0.5~m wide and was filled with fresh water to a depth $h = $~0.6~m. Waves were generated in one end of the tank with a computer controlled hydraulic piston wave maker. The perforated hose was centered on the bottom of the tank in the span-wise direction 11~m from the wave maker. Two powerful LED lamps were suspended from the tank walls and angled downwards to illuminate the bubbles from the sides. The ROV was placed outside the tank with the camera axis perpendicular to the bubble plane. The camera lens was located 30~cm from the tank wall, meaning that the distance between camera and bubbles was approximately 55~cm. The FOV in the tank center was approximately 80$\times$45~cm. Before the experiments were initiated, a Perspex plate marked with reference coordinates was placed vertically on the same location as the bubble plane and recorded with the camera. The reference coordinates, 88 in total, had a horizontal resolution of 5~cm and a vertical resolution of 2~cm close to the surface and 5~cm further down in the depth. Pixel coordinates were converted to real world coordinates with a cubic transform \citep{kolaas2016getting}. Fluid motion was analyzed within a section of 45~cm in the span-wise direction, located approximately in the center of the camera FOV, and 30~cm in the vertical direction from the undisturbed surface and downwards. This section was well within the domain of the pixel-to-world transformation. The distortion effect of the camera, which is most prominent in the outer edges, was also reduced in the center of the FOV.

\begin{figure}
  \begin{center}
    \includegraphics[width=.55\textwidth]{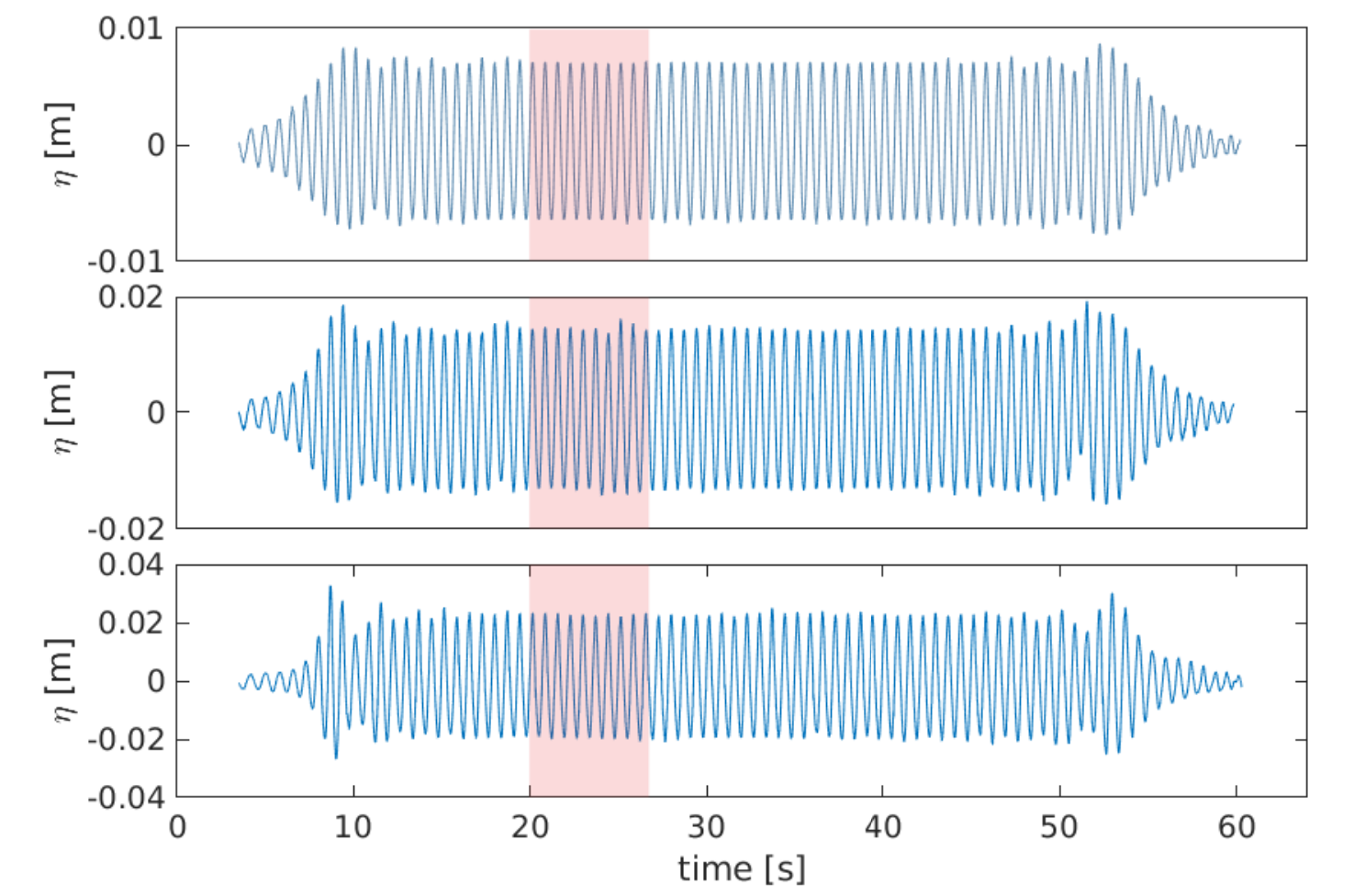}
    \caption{\label{fig:time_series} Time series of surface elevation $\eta$ obtained with ultrasonic gauges for 1.4~Hz monochromatic waves. The amplitude was approximately 7, 15 and 23~mm for top, middle and bottom panel, respectively. The shaded area marks the 10 periods investigated, which were periodic, almost constant in amplitude and unaffected by any reflected waves from the beach.}
  \end{center}
\end{figure}

A series of about 70 monochromatic waves was generated in each run. After a short transient build-up, the wave train became periodic and this is the part included in the analysis. An absorbing beach damped the waves at the far end of the tank and the analysis was terminated before any reflected waves reached back to the FOV. Figure~\ref{fig:time_series} shows the surface elevation $\eta$ in the position of the ROV with respect to time for 1.4~Hz waves and three different amplitudes. The 10 investigated periods are highlighted. Surface elevation at the test section of the tank was measured with an array of three ULS Advanced Ultrasonic wave Gauges (UGs) from Ultralab. The UG situated directly above the ROV was mainly used, and the two others were applied as redundancy in case the first one failed, which happened in a few runs. The UGs sampled at 250~Hz and they have a technical resolution of 0.18~mm.  

Wave amplitude $a$ from each run was determined as the mean value of $\eta_{max}$ over the 10 periods investigated. Three different wave frequencies $f=$~1.4, 1.6 and 1.8~Hz, and three different amplitudes of approximately 7, 15 and 23~mm were investigated. Each combination was repeated three times, which means 27 runs in total. The wavenumber $k$ of each run was determined from Eq.~\ref{eq:Dispersion}, which was used to find the wavelength $\lambda$. For all frequencies, $\lambda$ (0.51$-$0.82~m) was comparable to $h$. Hence, the waves were considered to be deep-water waves. In most runs, the wave steepness $ak < $~0.2. Only the two runs where the highest amplitude and the two highest frequencies were combined yielded 0.2~$ < ak < $~0.24. \cite{grue2003kinematics} showed that the velocity profile beneath a wave crest is very close to Stokes third order theory for $ak = $~0.23, and still a good approximation for $ak < $~0.30. Therefore, it is reasonable to use Eq.~(\ref{eq:Velocity_liquid_u}) as an approximation of theoretical $u_{w}$. 


Only bubbles located in vertical columns (0.04$\lambda$ wide) directly below the wave crests were considered. In the following results, $u_{b}$ of the smallest tracked bubbles is presented in order to minimize the horizontal oscillating motion \citep{haberman1953experimental}. The bubble radius was estimated from the observed vertical velocity, based on the findings presented in Section~\ref{subsec:Bubbles}. The vertical velocities accepted were 0.15~$<v_{b}<$~0.25~m/s, which should correspond to bubble radius 0.33~$<R<$~0.47~mm according to the logarithmic curve fit applied in the left panel of Fig.~\ref{fig:diam_vel_oldtube}. This range of bubble radius is marginally smaller than the bubbles investigated in Section~\ref{subsec:Bubbles}, most likely due to a slightly different input pressure on the perforated hose. For the 1.4~Hz waves, each run contained on average 683 analyzed bubbles which satisfied the quality criteria (small bubbles located directly below the crest) during the 10 periods investigated. In the case of 1.4~Hz waves, the 10 investigated periods consisted of approximately 214 image frames, meaning that 3.2 relevant bubbles were analyzed on average per image frame. 

\begin{figure}
  \begin{center}
    \includegraphics[width=1.02\textwidth]{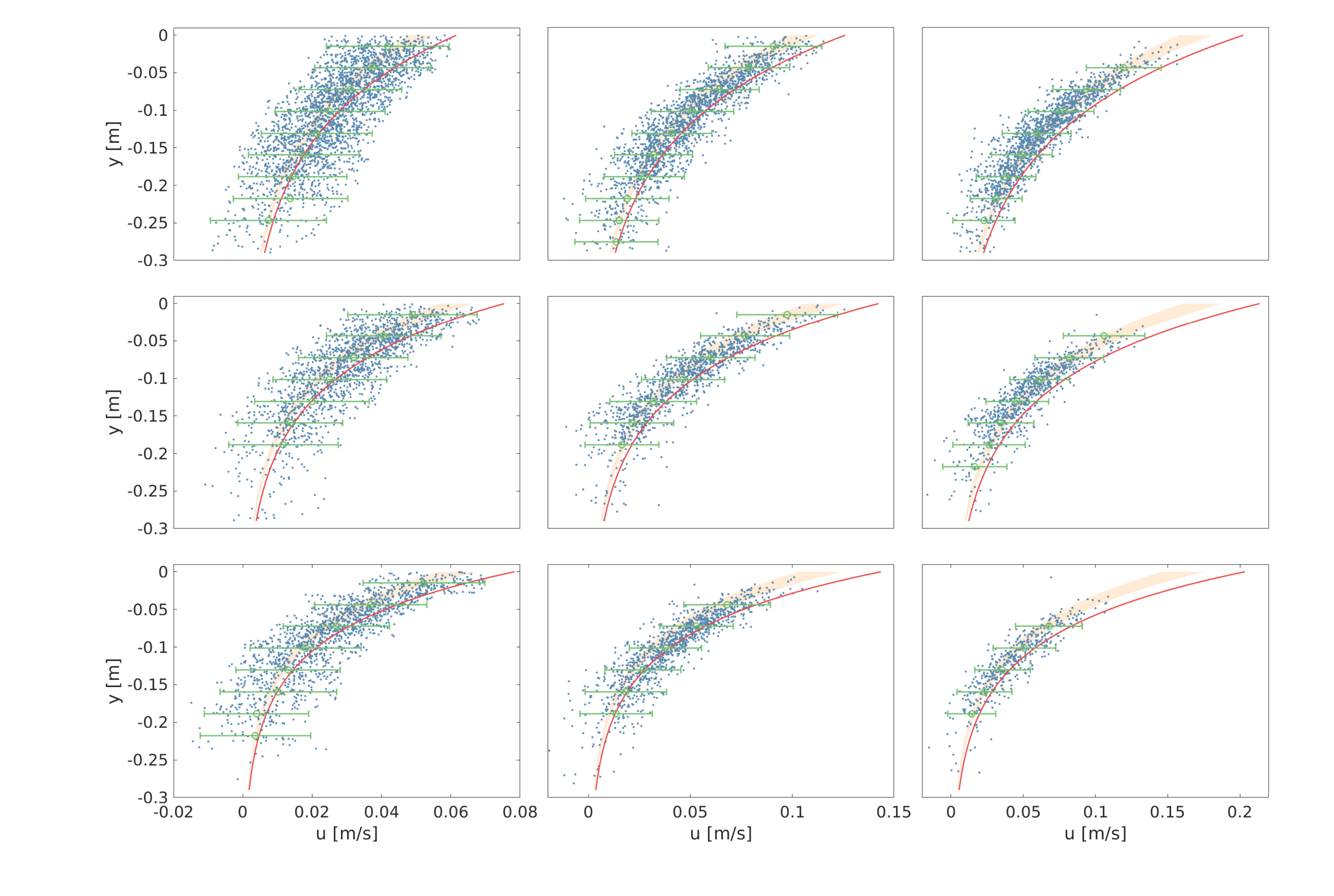}
    \caption{\label{fig:dim_profile} Horizontal velocity profiles divided into 10 equally spaced vertical bins below wave crests. Wave frequency: 1.4, 1.6 and 1.8~Hz from upper to lower row, respectively. Wave amplitude: approximately 7, 14 and 21~mm from left to right column, respectively. Blue dots: observed horizontal bubble velocity $u_{b}$. Green error bars: mean and $2 \sigma$ uncertainty of the observed $u_{b}$ within each bin (at least 10 observations must be present within the bin for the error bar to be displayed). Red line: theoretical horizontal water velocity $u_{w}$ given by Eq.~(\ref{eq:Velocity_liquid_u}). Orange shaded region: expected $u_{b}$ from estimated bubble size and Eq.~(\ref{eq:bubble_vel}). The observed mean $u_{b}$ lies within the expected region for most of the vertical bins. }
  \end{center}
\end{figure}

Figure~\ref{fig:dim_profile} shows the horizontal velocity profiles below wave crests with increasing frequencies from upper to lower panels and increasing amplitudes from left to right panels. All three repetitions of the same frequency-amplitude combination are included in each panel. The blue dots indicate observed $u_{b}$. Each profile is divided into 10 equally spaced vertical bins. A global filter was applied to remove outliers which were mostly situated around $y=$~0, probably caused by local surface tension effects or from bubbles accumulating on the surface. Horizontal velocities which deviated more than a certain threshold off the mean of the same bin were discarded. The threshold was set to 70\% of the total mean $u_{b}$ of all observed particles in the whole vertical profile. For the 1.4~Hz waves, this filter removed on average 2.9\% of the observations. After filtering, the mean and two standard deviations ($2\sigma$) of $u_{b}$ within each bin were calculated (green error bars). Each bin must contain at least 10 tracked velocities for the error bar to be shown. The theoretical fluid velocity given in Eq.~(\ref{eq:Velocity_liquid_u}) is displayed with a red line. The expected bubble velocity was calculated from Eq.~(\ref{eq:bubble_vel}), with the range of estimated bubble radius and the kinematic viscosity $\nu = $~10$^{-6}$~m$^{2}$/s inserted, and displayed as an orange highlighted region.

From Fig.~\ref{fig:dim_profile}, it can be seen that the horizontal bubble velocity in general has a smaller magnitude than the fluid velocity, but the exponential profile is clearly visible. The simple error estimate given in Section~\ref{subsec:Bubbles} and Eq. (\ref{eq:bubble_vel}) seems to be a good approximation for the slip velocity, as the observed mean lies within the expected bubble velocity for most of the vertical bins. The theoretical bubble velocity deviates 10.6$-$21.6, 12.2$-$24.7 and 13.7$-$27.7\% from the theoretical fluid velocity for the 1.4, 1.6 and 1.8~Hz waves respectively, depending on the bubble radius. Statistics from the second vertical bin from the top of all the 1.4~Hz waves are displayed in Table~\ref{table:dim_profiles}. This bin was chosen because the highest bin in terms of y-position and velocity does not contain enough tracked bubbles in the case of the highest wave amplitude. Here, the mean observed $u_{b}$ deviates with 1.3$-$6.5\% from the expected $u_{b}$ of the average estimated bubble radius. The spread in the observed data, exemplified through $\sigma$ in Table~\ref{table:dim_profiles}, slightly increases with increasing fluid velocity (i.e. higher wave amplitude). However, the relative standard deviation of the observations, that is $\sigma$ divided by mean observed $u_{b}$ in the respective bin, decreases from 22.4\% to 10.9\% from the lowest to the highest amplitude for the 1.4~Hz waves. Deeming from Fig.~\ref{fig:dim_profile}, the relative standard deviation of the observed velocities decrease in general with increasing fluid velocity (i.e. higher wave amplitude and/or frequency). The mean observed $u_{b}$ is 11.7$-$16.0\% smaller than theoretical $u_{w}$.        

\begin{table}[h]
\centering 
\begin{tabular}{c c c c c}  
\toprule
$a$ [mm] & Mean obs. $u_{b}$ [m/s] & $\sigma$ obs. $u_{b}$ [m/s] & PE $u_{w}$ [\%] & PE $u_{b}$ [\%] \\ [0.5ex]
\midrule
7 & 0.038 & 0.008 & -13.9 & 3.8 \\[0.5ex]
15 & 0.079 & 0.010 & -11.7 & 6.5 \\[0.5ex]
23 & 0.115 & 0.013 & -16.0 & 1.3 \\[0.5ex]
\bottomrule
\
\end{tabular} 
\caption{Statistics of observed $u_{b}$ for 1.4~Hz waves in the second vertical bin from the top. The fourth column from the left is the percentage error between observed $u_{b}$ and theoretical $u_{w}$ (from Eq.~\ref{eq:Velocity_liquid_u}) and the fifth column is the percentage error between observed and expected $u_{b}$ (from Eq.~\ref{eq:bubble_vel}). } 
\label{table:dim_profiles} 
\end{table}

\begin{figure}[htp]
\centering
\includegraphics[width=.32\textwidth]{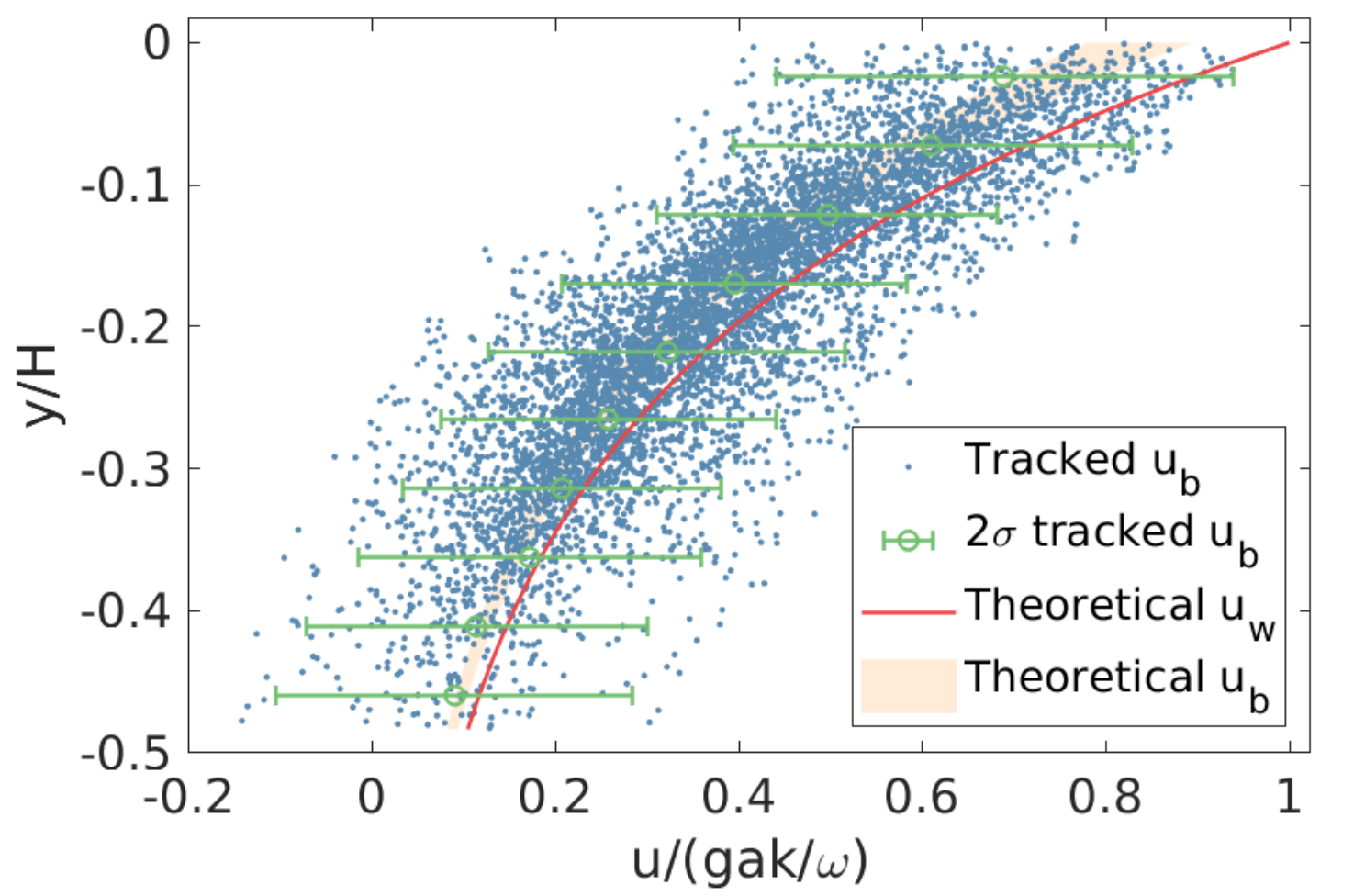}
\includegraphics[width=.32\textwidth]{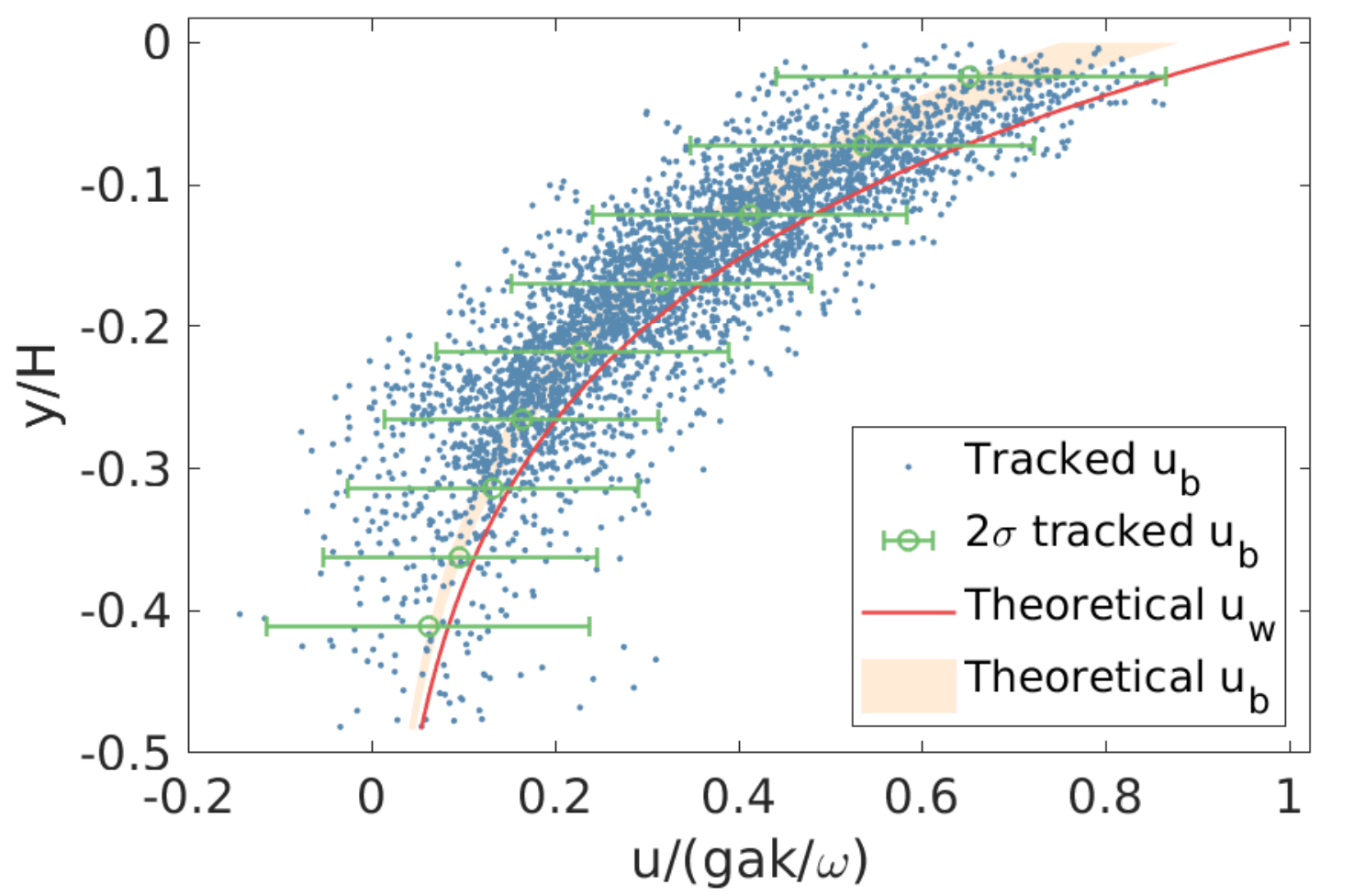}
\includegraphics[width=.32\textwidth]{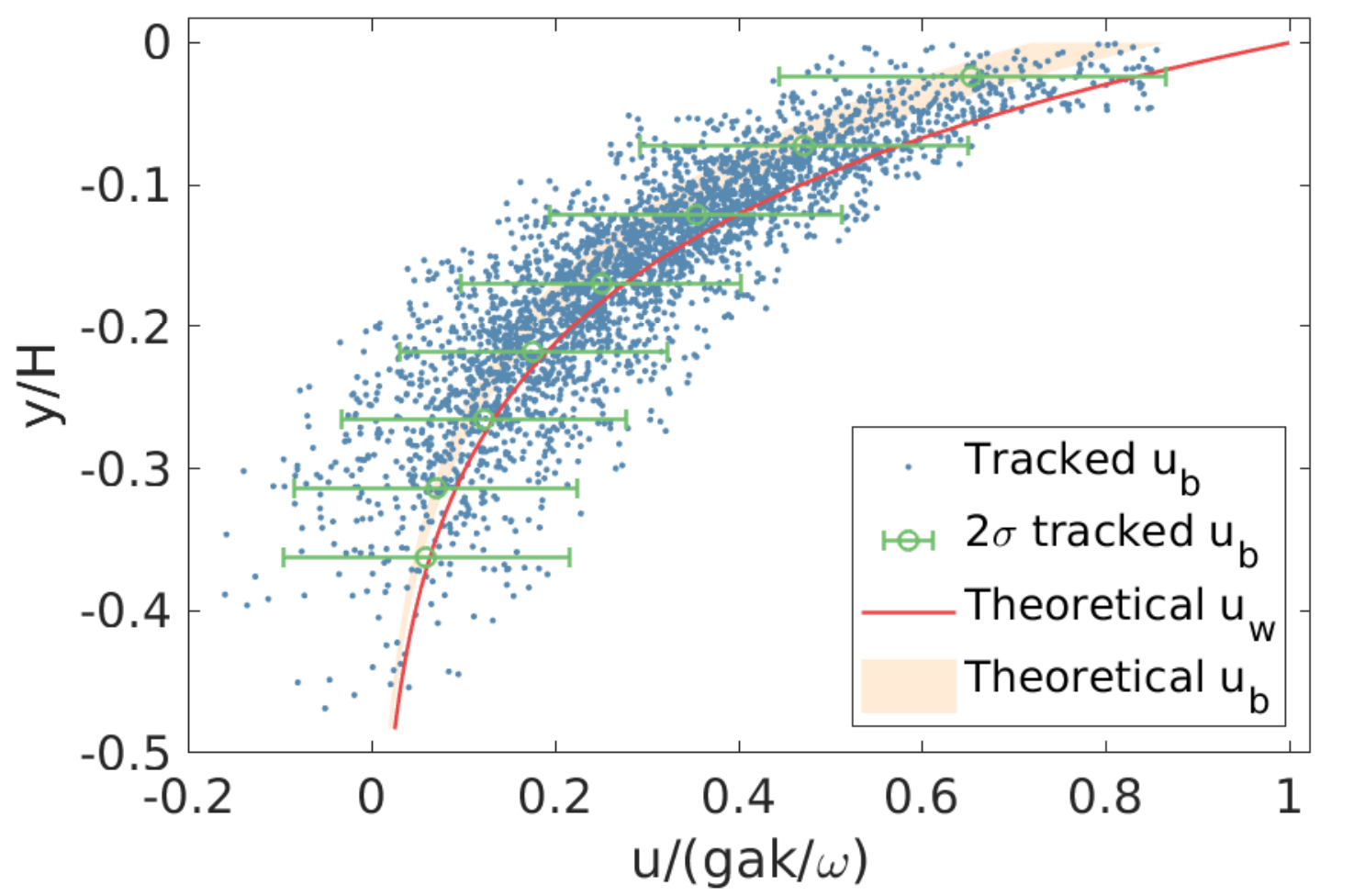}
\caption{Horizontal non-dimensional velocity profiles divided into 10 equally spaced vertical bins below wave crests. Wave frequency: 1.4, 1.6 and 1.8~Hz from left to right panel respectively. Blue dots: observed horizontal bubble velocity $u_{b}$. Green error bars: mean and $2 \sigma$ uncertainty of the observed $u_{b}$ within each bin (at least 30 observations must be present within the bin for the error bar to be displayed). Red line: $e^{ky}$. Orange shaded region: expected $u_{b}$ from estimated bubble size and Eq. (\ref{eq:bubble_vel}). The observed mean $u_{b}$ lies within the expected region for most of the vertical bins. }
\label{fig:nondim_profile}
\end{figure}

The velocity profiles were nondimensionalized and the different amplitudes with the same frequency are plotted together with increasing frequency from left to right panel in Fig.~\ref{fig:nondim_profile}. Outliers were filtered in the same manner as in Fig.~\ref{fig:dim_profile}. Here, the mean and error bar is presented if a bin contains at least 30 tracked bubbles. A clear collapse of the different amplitudes can be seen for each frequency. The mean of the observed $u_{b}$ is located within the expected region for most vertical bins. The wavenumbers from the different amplitudes are averaged to show theoretical $u_{w}$ and expected $u_{b}$ for each frequency. Visually, the spreading of tracked horizontal velocities seems to decrease with increasing frequency. This observation is consistent with the decreasing relative standard deviation for higher velocities, which was the case for the dimensional graphs in Fig.~\ref{fig:dim_profile}.     

\section{Field deployment} \label{sec:Field_deployment}

\begin{figure}
  \begin{center}
    \includegraphics[width=.45\textwidth]{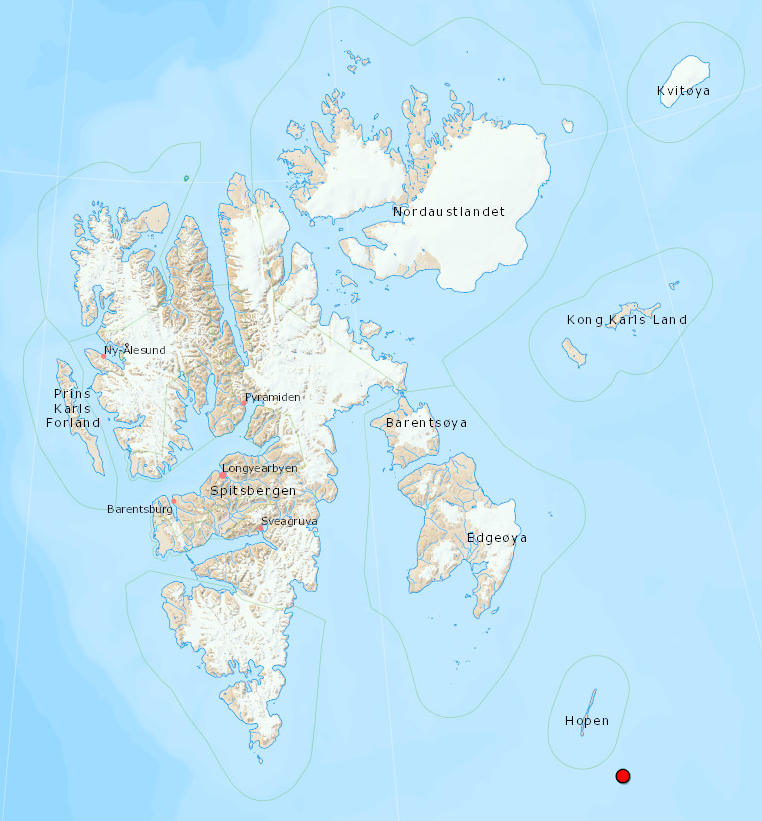}
    \caption{\label{fig:map} A map of the area where the field work was carried out. The red dot indicates the location of the ice floe. Source: \cite{TopoSvalbard}.}
  \end{center}
\end{figure}

\begin{figure}[!h]
       \centering
       \includegraphics[width=0.49\linewidth,height = 5.5cm]{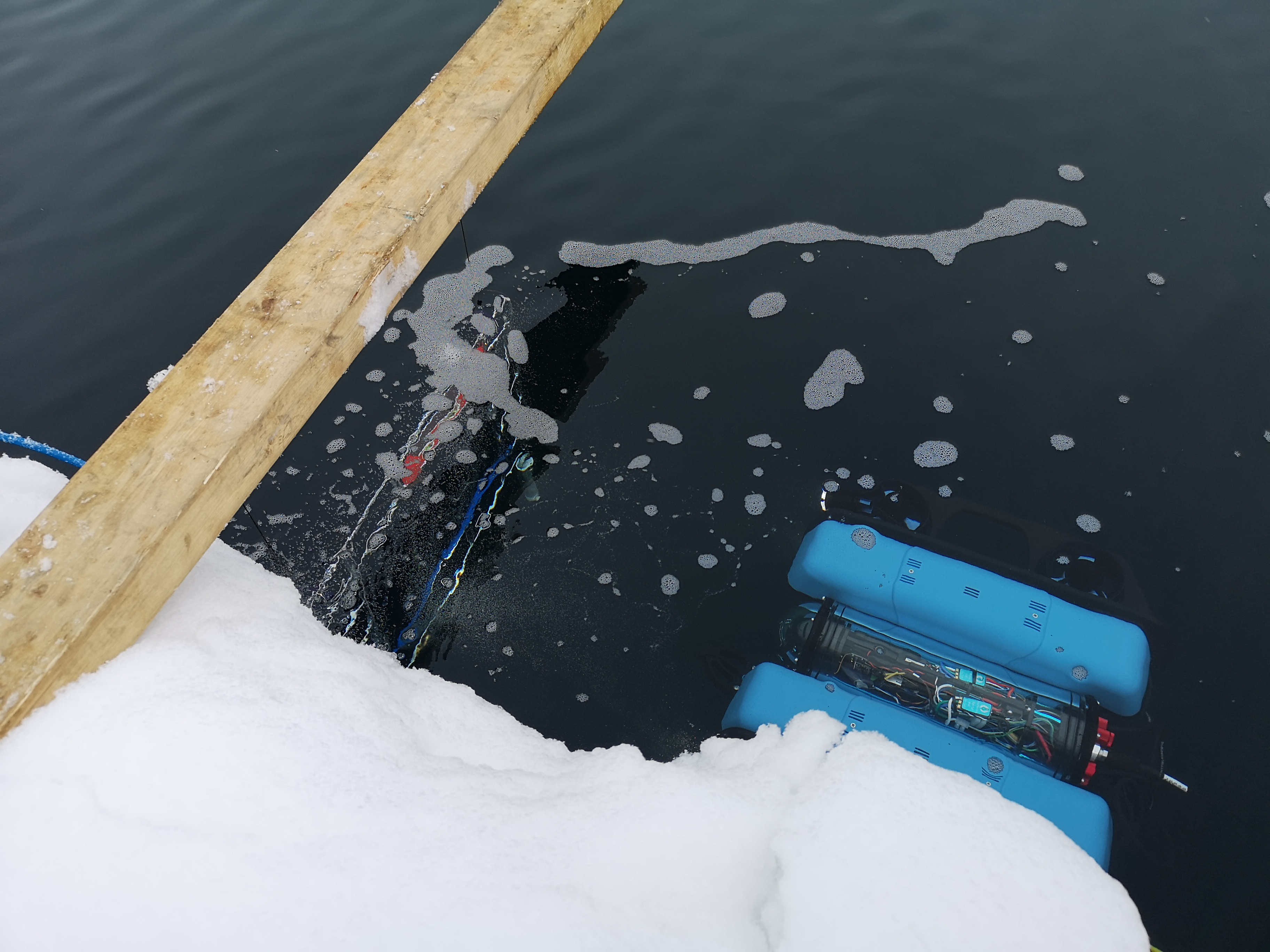}
       \includegraphics[width=0.49\linewidth,height = 5.5cm]{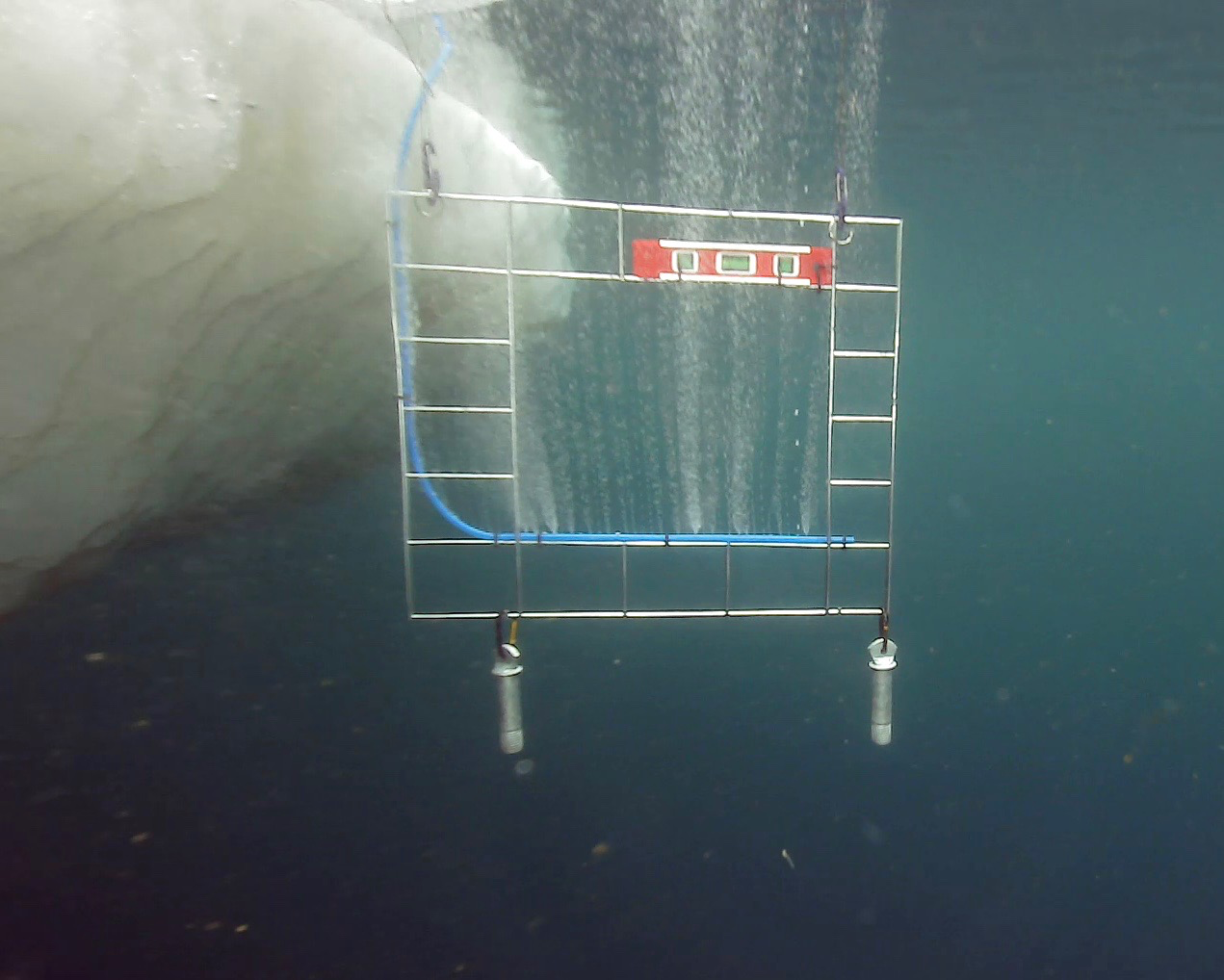}
      \caption{ The ROV located in front of the bubble plane and the coordinate grid, which were lowered into the water by the pulley system, seen from above (left panel) and from the ROV perspective (right panel). A spirit level was used to align the coordinate grid with the horizontal. }
      \label{infront_under}
\end{figure}

The ROV-PV system was tested close to an ice floe in the North-West Barents Sea on April 26, 2019. The objectives of the field experiment were to test the setup under Arctic conditions and to investigate the flow around floating ice subjected to wave motion. The site was near Hopen Island, which is a part of the Svalbard Archipelago shown in Fig.~\ref{fig:map}, and the geographical coordinates were 76.18$^{\circ}$N, 25.77$^{\circ}$E (indicated by the red dot). The setup was lowered down from a beam by a pulley system into the water next to the floe, which had a diameter of approximately 20~m. Figure~\ref{infront_under} shows the ROV positioned in front of the grid (left panel) and the bubble plane seen from the ROV (right panel). The bubbles were illuminated by both ambient light and the ROV headlights. 

Due to the dense bubble plane and the relatively long distance from the ROV to the bubbles, it proved difficult to track single bubbles between frames obtained next to the ice floe. Therefore, PIV was applied to obtain a 2D velocity field. The PIV processing was performed with subwindows of 48$\times$48 pixels and 75$\%$ overlap. A search range of 1/3 of the subwindow was used. This configuration yielded approximately 20$\times$16 2D velocity vectors inside the bubble plane. Each velocity vector can be decomposed into $u_{b}$ and $v_{b}$, i.e. a horizontal and a vertical component, respectively. 

In the attempt of finding $v_{w}$, two different approaches have been tested to compensate for the vertical buoyancy driven velocity component of the bubbles; by subtracting either the most frequently observed terminal velocity in the laboratory experiment with salt water, or the mean vertical velocity observed in the field. In the first case, the velocities observed in the right panel of Fig.~\ref{fig:diam_vel_oldtube} are distributed into bins with 0.02~m/s resolution. The mode of the vertical velocity, $\overline{v_{b,lab}}=$~0.37~m/s, is subtracted from $v_{b}$, i.e. $v_{w} = v_{b} - \overline{v_{b,lab}}$. In the second case, a reference image couple from the field experiment where the horizontal velocity component is very small, is chosen. From the PIV analysis of this image couple, the mean velocity components $\overline{u_{b,field}}=$~0.0058~m/s and $\overline{v_{b,field}}=$~0.15~m/s are found by spatially averaging the horizontal and vertical velocity components, respectively. Then, $\overline{v_{b,field}}$ from the reference image couple is subtracted from all $v_{b}$ in the image couple that is analyzed, i.e. $v_{w} = v_{b} - \overline{v_{b,field}}$. In the latter case, it is assumed that $v_{w}$ is small at the time of the reference image couple, since the observed $u_{w}$ is small. 

\begin{figure}
    \includegraphics[width=0.33\linewidth,height = 4.5cm]{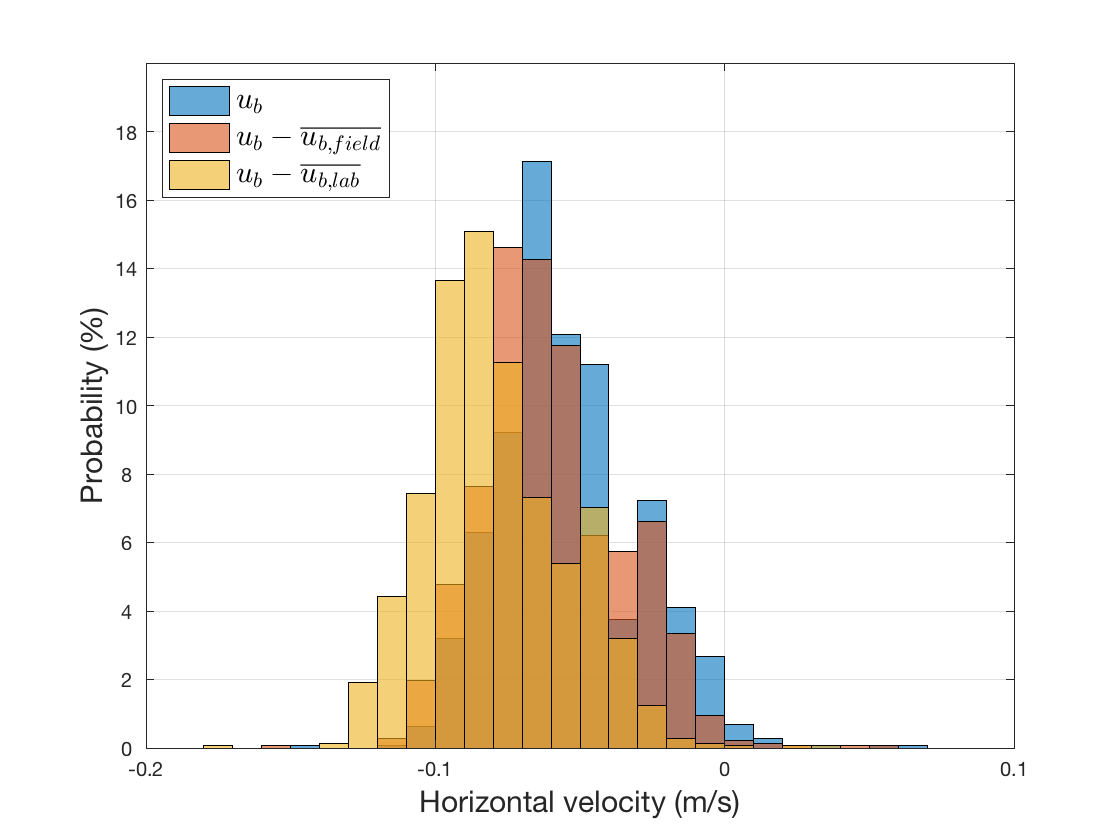}
    \includegraphics[width=0.33\linewidth,height = 4.5cm]{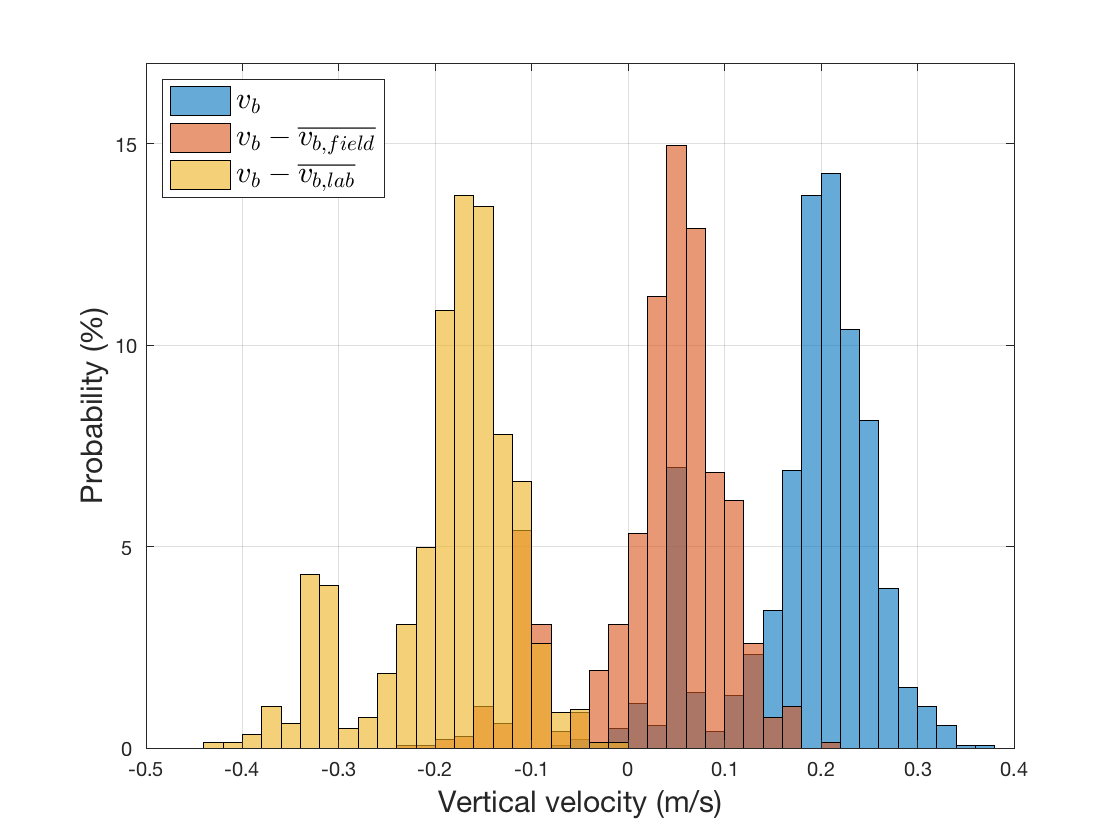}
    \includegraphics[width=0.33\linewidth,height = 4.5cm]{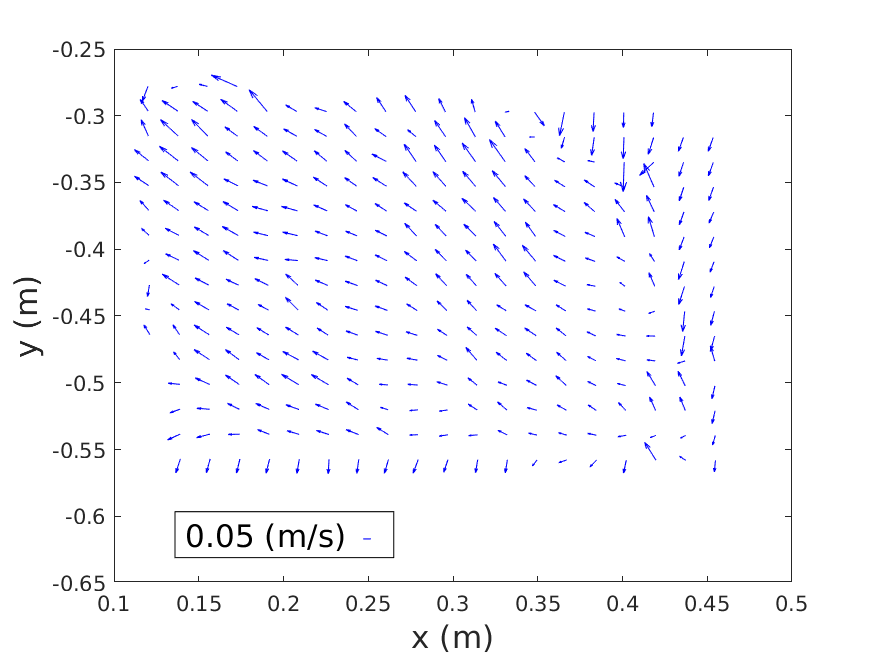}
  \caption{ Example of an instantaneous velocity field obtained next to the ice floe. Probability distributions of the tracked bubble velocity (blue), and estimates for the water velocity from laboratory (yellow) and reference field (orange) experiments in the horizontal (left panel) and vertical (center panel) direction. Quiver plot estimated from field reference experiment (right panel). }
  \label{fig:quiver}
\end{figure}

An example from an instantaneous velocity field obtained next to the ice floe is presented in Figure~\ref{fig:quiver}. Measured bubble velocities are distributed into bins with 0.02~m/s resolution and displayed (blue) in the left ($u_{b}$) and center ($v_{b}$) panels. The probability distribution for the vertical water velocity obtained from the two different approaches, i.e. $v_{b} - \overline{v_{b,field}}$ (orange) and $v_{b} - \overline{v_{b,lab}}$ (yellow), are shown in the center panel. The same procedure is followed in the horizontal direction in the left panel, although this is only for illustrative purpose, as we assume that $u_{b} \approx u_{w}$. The two different approaches yield quite different results in the vertical direction. The terminal velocity observed in the laboratory was in general more than twice as high as in the field. A possible explanation for this deviation is the temperature difference, which may have changed the flexibility of the perforated hose. For example, the diameter of the holes where the bubbles escape may decrease in lower temperature, or they could be partly clogged by ice. The most realistic results are obtained with $v_{b} - \overline{v_{b,field}}$ because the conditions in the field differ so much from the laboratory. Additionally, this approach gives similar water velocity magnitudes in the horizontal and vertical direction.  

A quiver plot is presented in the right panel of Fig.~\ref{fig:quiver}. It is produced with the latter approach, i.e. subtraction of the mean reference velocity from the field, in both horizontal and vertical direction. The undisturbed ocean surface is set to $y = $~0 and the ice floe is located at a negative x-value. The direction of the water flow is towards the ice floe. In fact, it was observed from the ROV images that the horizontal velocity oscillated towards and away from the ice floe with periods in the order of 10~s. This motion is either direct wave motion or flow generated from the heaving ice floe. There may also exist a circulating structure covering about 30\% of the right side of the quiver plot, which could be a turbulent eddy generated by the ice floe.

\section{Discussion} \label{sec:Discussion}

In order to determine the accuracy of the introduced ROV-PV system, potential sources of error must be identified and quantified. Fluid velocity is estimated from the motion of bubble tracers by identifying the bubble position with an image processing technique. Two central questions that need to be addressed are how accurately the bubble position can be determined and how well a particle follows the fluid motion. According to \cite{dalziel1992decay}, the position of particles spanning over one pixel is usually determined with pixel accuracy, while the uncertainty decreases for particles spanning over several pixels. In the wave tank experiments, the bubbles typically covered at least a couple of pixels. Therefore, one pixel is used as a conservative accuracy estimate of the bubble position, which corresponds to approximately 0.4~mm in real world coordinates. The bubble velocity is determined from a central differencing scheme, i.e. over three image frames. With a typical vertical velocity of 0.25~m/s, a bubble travels 0.025~m during three frames. This gives 1.6\% relative error in velocity.

When it comes to the passivity of the particles or their capability of following the motion of the surrounding fluid, it has been shown that the horizontal bubble velocity lags behind the fluid velocity under periodic waves. This effect is attributed to the fluid acceleration and the bubble inertia due to the effect of added mass. For the 1.4~Hz waves, the discrepancies between observed bubble velocity and theoretical water velocity were 11.7$-$16\%. The relative error in velocity caused by the bubble inertia increases with the bubble radius and is theoretically found to be 10.6$-$21.6\% for the bubbles investigated under the 1.4~Hz waves. Hence, the error due to the bubble inertia is an order of magnitude larger than the error due to the image processing technique. The estimated horizontal slip velocity agrees with the observations. Note that the observed horizontal slip velocity yields Reynolds numbers of unity order of magnitude, which suggests that Stokes drag is a reasonable approximation.

A considerable spread in the observed $u_{b}$ under the wave crests can be seen from Figs.~\ref{fig:dim_profile}-\ref{fig:nondim_profile}, where $\sigma \approx $~0.01~m/s. This spread is partly related to the varying bubble size and therefore varying slip velocity, as indicated by the orange shaded regions in the figures, but also due to secondary motion, typically oscillation of rising bubbles. The perforated hose was designed to produce as small bubbles as possible to obtain rectilinear trajectories. Although this was achieved to a certain extent, Fig.~\ref{fig:diam_vel_oldtube} shows maximum absolute horizontal velocity values of approximately 0.03~m/s for terminal velocity around 0.25~m/s ($Re\approx$~300). This observation is in agreement with \cite{clift2005bubbles}, who described onset of secondary motion from $Re=$~200$-$1000, where the typical horizontal component is 5\% of the vertical component. Oscillations are most likely caused by vortex shedding. The magnitude of the horizontal velocity component of a rising bubble can be considered analogous to the oscillating part of surface waves, see e.g. Eq.~(\ref{eq:Velocity_liquid_u}). Its magnitude will depend on y-position and time. For bubbles of the same size, we can expect a uniform distribution in the horizontal velocity component due to oscillations. Since the bubble size varies, the distribution should be denser around the mean, which appears to be the case by visual inspections of Fig.~\ref{fig:nondim_profile}.


There is a large uncertainty associated with measurements of the vertical water velocity component since the bubbles are naturally buoyant. Bubbles rising in calm water and vertical columns directly below wave crests have been investigated in laboratory experiments. In both cases, the vertical water velocity is known to be zero, so that the vertical bubble velocity equals the vertical slip velocity. However, such controlled environment is seldom obtained in the field. For accurate measurements of the vertical velocity component, it is necessary to generate uniform bubbles in size. In this way, the terminal velocity due to the buoyancy can be found in advance and assumed within reasonable accuracy to be constant for all bubbles. The vertical fluid velocity can then be found from Eq.~(\ref{eq:Relative_vel}) in the case of steady flow, or from a combination of Eq.~(\ref{eq:Relative_vel})-(\ref{eq:Momentum_balance}) in the case of accelerated flow. 

The current ROV-PV system with bubbles as tracing particles is suitable for measuring flows where the horizontal velocity component is of interest. Examples of such situations are waves propagating under an ice cover and horizontal shear flow in the boundary layer beneath an ice cover. As discussed in the previous paragraph, a uniform bubble size is necessary to accurately measure the vertical velocity component. Further effort must be spent in order to improve the perforated hose to overcome this challenge in future studies. At the same time, focus should be directed towards keeping the construction simple in use and robust for field conditions. Hydrogen bubbles generated from hydrolysis could be an option as these tend to be small (diameter around 0.1~mm), although \cite{weier2013two} reported a distribution in bubble size due to coalescence phenomena. On the other hand, bubbles cannot be identified by the particle tracking software if they are too small. PIV might be suitable in this case.

\section{Conclusions} \label{sec:Conclusions}

In this paper, a new method for measuring 2D velocity fields in the upper ocean by utilizing image processing technology and air bubbles as tracing particles has been presented. The novelty of this method is the combination of bubbles and an ROV, which gives a simple and lightweight setup which is suitable for field measurements in the polar regions. The ROV-PV system has been demonstrated to measure the flow in the vicinity of an ice floe during an Arctic field campaign. There is a need for detailed ocean kinematics in the marginal ice zone, as a lot of important atmosphere-ice-ocean energy transfer processes occur here and relatively few in situ observations of 2D velocity fields exist. The introduced technique could be suitable for visualization and quantification of many interesting flow phenomena in the Arctic, for example wave propagation underneath various ice layers or turbulent structures induced by colliding ice floes. Detailed observations within this field are important for an improved understanding of the underlying physics and the design and validation of numerical sea ice models.

Detailed laboratory experiments have been carried out to quantify the relation between diameter and terminal velocity of the generated bubbles. Although many approximations and models exist for this relation, the present results demonstrate the importance of a thorough investigation of bubble properties since substantial deviations to models may occur. The ROV-PV system has been utilized to measure horizontal velocities under periodic water waves with an accuracy in the order of 10\%. The deviations are mainly attributed to the slip velocity caused by the bubble inertia due to the added mass, and is important to consider when measuring accelerated flow. In addition, there is a spread in the data expressed through a relative standard deviation in the order of 10\%, due to the bubble size distribution and the oscillatory motion of rising bubbles. Future studies should focus on generating smaller and more uniform bubbles, which could decrease the spread in observed velocities and improve the reliability in measurements of the vertical velocity component.

\section*{Acknowledgement}

The authors are grateful to Aleksey Marchenko for inviting us to the Hopen cruise. We also thank the crew of Polarsyssel for their assistance. Funding for the experiment was provided by the Research Council of Norway under the PETROMAKS2 scheme (project DOFI, Grant number $28062$). The data are available from the corresponding author upon request. 

\bibliographystyle{agsm}
\bibliography{template}

\end{document}